\documentclass[%
 reprint,
 prb,
]{revtex4-2}
\usepackage{graphicx}
\usepackage{dcolumn}
\usepackage{bm}
\usepackage{float}
\usepackage[colorlinks=true,linkcolor=blue,anchorcolor=red,citecolor=blue,urlcolor=black]{hyperref}
\usepackage[mathlines]{lineno}
\usepackage{ulem}
\usepackage{amssymb}
\usepackage{booktabs}
\begin{document}
\title{Field-Effect Josephson Diode \\via Asymmetric Spin-Momentum-Locking States}
\author{Pei-Hao Fu$^{1,2,3}$}
\email{peihao\_fu@sutd.edu.sg}
\author{Yong Xu$^{1,5}$}
\author{Shengyuan A. Yang$^{2,4}$}
\author{Ching Hua Lee$^{3}$}
\email{phylch@nus.edu.sg}
\author{Yee Sin Ang$^{2}$}
\email{yeesin\_ang@sutd.edu.sg}
\author{Jun-Feng Liu$^{1}$}
\email{phjfliu@gzhu.edu.cn}
\affiliation{$^1$School of Physics and Materials Science, Guangzhou University, Guangzhou 510006, China}
\affiliation{$^2$Science, Mathematics and Technology, Singapore University of Technology and Design, Singapore 487372, Singapore}
\affiliation{$^3$Department of Physics, National University of Singapore, Singapore 117542}
\affiliation{$^4$Research Laboratory for Quantum Materials, Singapore University of Technology and Design, Singapore 487372, Singapore}
\affiliation{$^5$Institute of Materials, Ningbo University of Technology, Ningbo 315016, China}
\begin{abstract}
Recent breakthroughs in Josephson diodes dangle the possibility of extending conventional non-reciprocal electronics into the realm of superconductivity.
While a strong magnetic field is recognized for enhancing diode efficiency, it concurrently poses a risk of undermining the essential superconductivity required for non-dissipative devices.
To circumvent the need for magnetic-based tuning, we propose a field-effect Josephson diode based on the electrostatic gate control of finite momentum Cooper pairs in asymmetric spin-momentum-locking states. 
We propose two possible implementations of our gate-controlled mechanism: (i) a topological field-effect Josephson diode in time-reversal-broken quantum spin Hall insulators; and (ii) semiconductor-based field-effect Josephson diodes attainable in current experimental setups involving a Zeeman field and spin-orbit coupling. 
Notably, the diode efficiency is highly enhanced in the topological field-effect Josephson diode because the current carried by the asymmetric helical edge states is topologically protected and can be tuned by local gates.
In the proposed Josephson diode, the combination of gates and asymmetric spin-momentum-locking nature is equivalent to that of a magnetic field, thus providing an alternative electrical operation in designing nonreciprocal superconducting devices.
\end{abstract}
\maketitle

\section{Introduction}
\begin{figure*}
\centering \includegraphics[width=1\textwidth]{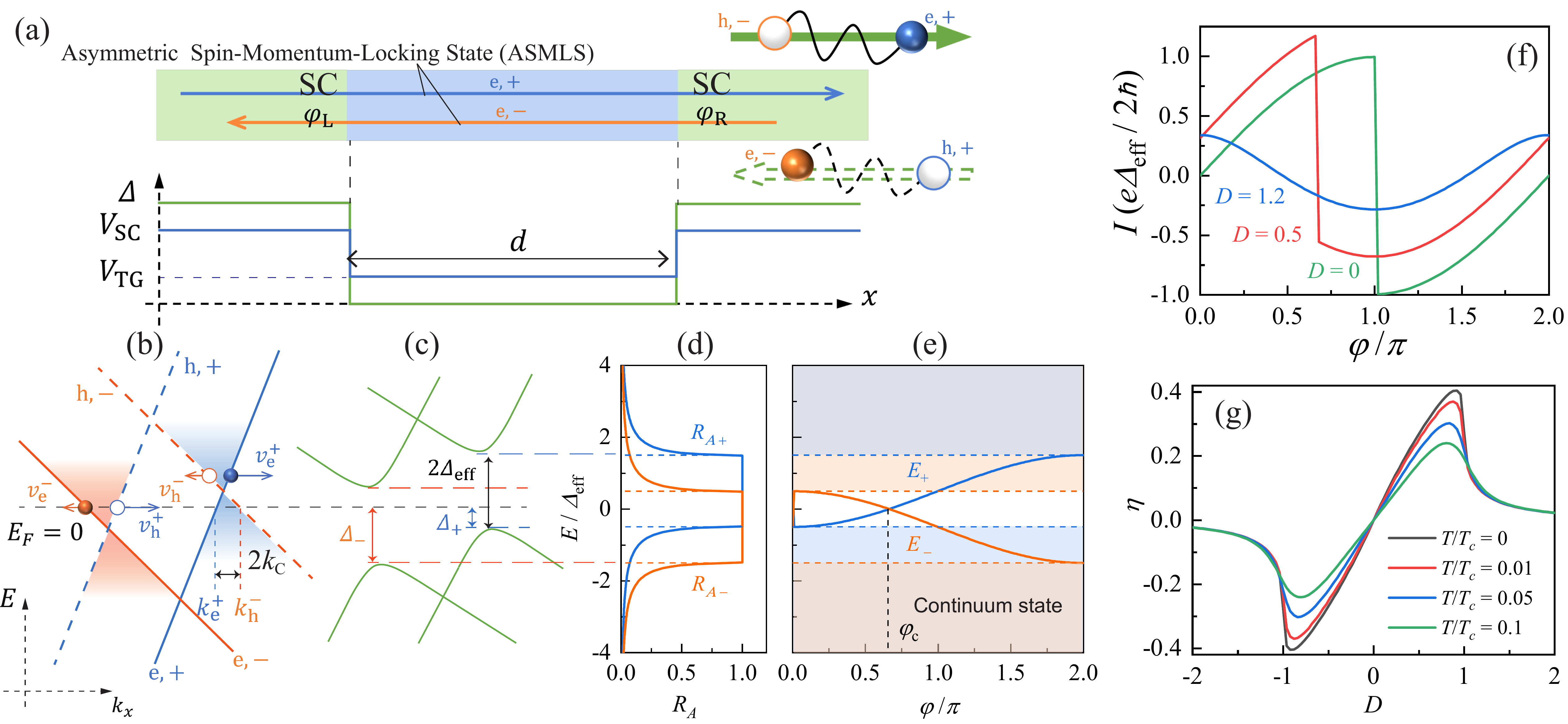}
\caption{
Mechanism of FEJD mediated by ASMLSs. 
(a) FEJD schematics, where the rightward-flowing dissipationless (leftward-flowing dissipative) current, labeled by solid (dashed) green arrow, is carried by Cooper pairs (unpaired carriers) formed by spin-up (spin-down) electrons and spin-down (spin-up) holes represented by blue (orange) solid and hollow circles, respectively. 
FEJD is controlled by super gates ($V_{\text{SC}}$) and tunneling gate ($V_{\text{TG}}$) applied to superconducting (SC) leads with an order parameter $\Delta$ and central normal region with a length $d$, respectively. 
(b) and (c): ASMLS dispersion, (b) without, and (c) with $\Delta$.
In (b), solid (dashed) blue/orange lines represent spin-up/down electrons (holes) dispersion. 
Velocity of spin-up (spin-down) quasiparticles are $v_{e/h}^{+(-)}$ 
as denoted by the length of the arrows. 
FEJD occurs due to the gate-tunable center-of-mass momentum of paired electrons $k_{\text{C}}$ [Eq. (\ref{kc})] that causes Doppler energy shift $D$ [\ref{DES}] and spin-dependent SC gaps below $E_F=0$ as $\Delta_{\pm }=\Delta _{\text{eff}}( 1\mp D)$ in (c). 
(d-g) are transport characteristics of FEJD with $\hbar v_{0}=1=\Delta $, $t=0.1$, and $D=0.5$.
(d) Asymmetric energy-resolved Andreev reflection probabilities [Eq. (\ref{ra})], 
i.e. $R_{A,\pm}( E) \neq R_{A,\pm }(-E) $. 
where the unit-probability energy regime corresponds to ABS spectra [Eq. (\ref{eabs1})] in (e) with $\varphi _{c}=2\cos ^{-1}D$. 
ABS accompanied by continuum states contribute to supercurrent [Eq. (\ref{i1})] in (f). 
The latter solely depends on $D$ resulting in a gate-tunable diode efficiency [Eq. (\ref{eta1})] in (g).}

\label{setup}
\end{figure*}

Electronic technologies rely heavily on non-reciprocal charge-transport semiconducting devices such as diodes, alternating-to-direct current converters, and transistors \cite{Fruchart2021,Zhang2023,Zhao2023,Stegmaier2023}. 
However, dissipation is inevitable in such devices due to their finite resistance \cite{Sze2012}.
Superconducting (SC) diodes \cite{Nadeem2023,Wakatsuki2017} in noncentrosymmetric superconductor \cite{Zhang2020,Ando2020,Schumann2020,Bauriedl2022,Itahashi2020,Shin2021,Lyu2021,Yasuda2019,yWu2022,JMasuko2022,Lin2022,Narita2022,Hou2022,Jiang2022}
and Josephson junctions \cite{Bocquillon2017,DiezMerida2021,Portoles2022,Turini2022,Gupta2022,Pal2022,Baumgartner2022a,Baumgartner2022b,Jeon2022,Golod2022,Wu2022,Ghosh2022,Chiles2022,Mazur2022,Anwar2022, Costa2022,FZhang2023,Song2022,Trahms2022}  
are promising dissipationless non-reciprocal devices that manifest in direction-selective critical supercurrents, i.e. $I_{c+} \neq I_{c-}$ where $I_{c+/-}$ is the critical supercurrent in positive/negative direction. 
When an external applied current lies between $I_{c-}$ and $I_{c+}$, the transport is dissipationless in one direction due to Cooper-pair formation, while dissipative in the opposite direction due to the destruction of Cooper pairs [Fig. \ref{setup}(a)]. 

The extent of non-reciprocity is characterized by the diode efficiency,
\begin{equation}
\eta =\frac{I_{c+}-I_{c-}}{I_{c+}+I_{c-}}\text{,}  \label{etadef}
\end{equation}
and $sign(\eta)$ defines the diode polarity. 
Numerous theoretical findings \cite{Hoshino2018,Wakatsuki2018,Zhai2022,He2022,Scammell2022,Ili2022,Yuan2022,Daido2022a,Legg2022,Daido2022b,Chen2018,Kopasov2021,Misaki2021,Halterman2022,Wei2022,Tanaka2022,Zhang2022,Soori2022a,Soori2022b,Davydova2022,Souto2022,Fominov2022,Lu2022,Hu2022,Haenel2022,Steiner2022,Kokkeler2022,Tanaka2022,Maiani2023,Legg2023,Xie2022,Ding2022,Ding2023,Alidoust2021,Karabassov2024} in topological insulators \cite{Legg2022,Chen2018,Lu2022,Kokkeler2022,Tanaka2022,Legg2023,Karabassov2024}, half-metallic trilayers \cite{Halterman2022}, supercurrent interferometers \cite{Souto2022,Fominov2022} and, semiconductors subjected to the interplay with spin-orbit interactions (SOIs) and Zeeman field \cite{Maiani2023,Alidoust2021} have suggested SC diodes as potential building blocks for high-performance SC circuits.


Simultaneously breaking inversion and time-reversal symmetry (TRS) is necessary for SC diode effect \cite{Zinkl2022,Hu2007,Wang2022}. 
A resulting additional unidirectional supercurrent is induced causing the non-reciprocity \cite{Davydova2022,Yuan2018,Yuan2021,Hart2017}.
To break inversion symmetry, a nuanced device design is essential where the diode efficiency is sensitive to structural parameters \cite{Narita2022}, and challenging to tune once a device is fabricated.
Thus, in most experiments, SC diode effects are activated and manipulated by external magnetic fields, a typical method for breaking TRS.
Unfortunately, superconductivity will ultimately be impaired by sufficiently high magnetic fields, thereby fundamentally limiting their applications \cite{Muller2003}. 
Attention then is drawn to magnetic-field-free SC diodes \cite{Lin2022,Jiang2022,Narita2022,Golod2022,FZhang2023,Jeon2022,Wu2022,Mazur2022}, where TRS is broken by intrinsic magnetism. 
However, external magnetic fields are still necessary for diode-on/off and polarity-switching \cite{Jeon2022}. 
As such, an all-electrically-controlled SC diode is still under demand, which enhances the flexibility in operating integrated SC circuits.

This work proposes a new mechanism for an all-electrically-controlled Josephson diode, wherein supercurrents are carried by asymmetric spin-momentum-locking states (ASMLSs) characterized by electrons with opposite spins counter-propagating with different velocities (Fig. \ref{setup}).
Specifically, we show that the combination of asymmetry and gate voltage in the SC leads plays an equivalent role as a magnetic field to control the center-of-mass velocity of paired electrons, which leads to a unidirectional supercurrent.
This enables the construction of an all-electrically controllable Josephson diode circumventing magnetic control.
Hence, the proposed device can be termed as a \textit{field-effect Josephson diode} (FEJD).

A proof-of-concept FEJD is proposed in a specific type of TRS-broken quantum spin Hall insulator (QSHI) with asymmetric helical edge states, which is predicted in Floquet topological insulator \cite{Fu2022} and magnetic-doped QSHI with a small magnetic field \cite{ Chen2021, Zhang2019} such as HgTe \cite{Hart2017,Bocquillon2017} and MnBi$_2$Te$_4$ \cite{Chen2022,WZXu2022}.
Since the supercurrents carried by the asymmetric helical edge states are topologically protected \cite{Tkachov2015, Scharf2021, Dolcini2015, Alidoust2017, Beenakker2013, Tkachov2019a,Tkachov2019b,Vigliotti2023}, resembling the topological Josephson junction \cite{Dolcini2015}, the resulting device can be dubbed as \textit{topological} FEJD.

The mechanism of realizing a FEJD using ASMLSs is also attainable in semiconductors which are widely investigated in recent experiments \cite{Jeon2022,Baumgartner2022a,Baumgartner2022b,Costa2022,Gupta2022,Turini2022,Mazur2022}. 
With the interplay between spin-orbit interaction (SOI) and in-plane Zeeman fields, electrons near the Fermi surface of the sample are described by ASMLSs. 
Our mechanism attributing the gate-tunable behavior to the ASMLSs provides an alternative theoretical explanation on a recent experiment \cite{Mazur2022}.

The remainder of this work is organized as follows. 
The properties of ASMLSs and the mechanism of FEJD induced by ASMLSs are introduced in Sec. \ref{sec_kc}.
The mechanism is schematically demonstrated in Fig. \ref{setup}, where the gate-tunable center-of-mass momentum of paired electrons and the Doppler energy shift play critical roles. 
The conceptual FEJD is specialized in two scenarios: topological FEJD in QSHI with asymmetric helical edge states (Sec. \ref{sec_topfejd}) and semiconducting systems subjected to the interplay between SOIs and Zeeman fields (Sec. \ref{sec_semifejd}).
The results are supported by the details involving scattering matrix formalism and tight-binding model simulation in the appendix \ref{App_A}-\ref{App_E}.
Conclusions and discussions on experimental realization and the universality of ASMLS are given in Sec. \ref{sec_discon}.


\section{General behavior of field-effect Josepshon diode} \label{sec_kc}


The mechanism of FEJD induced by ASMLSs is demonstrated in this section schematized in Fig. \ref{setup}.
The general properties of ASMLS are introduced in Sec. \ref{sec_kc_asmls}, including its characteristics and potential implementation to the gate-tunable center-of-mass momentum of paired electrons when SC pairing is considered.  
The general transport properties of FEJD are demonstrated in Sec. \ref{sec_kc_genfor} including the asymmetric Andreev reflection probabilities, Andreev bound states (ABSs) spectra, current-phase relation (CPR) which are all-electrically controllable, The resulting gate-controlled diode efficiency is up to $40\%$ which is comparable to the theoretical value obtained by applying a magnetic field. 


\subsection{Asymmetric spin-momentum locking states} \label{sec_kc_asmls}


To begin with, we employ a representative 2-component ASMLS Hamiltonian $h_{0}$, with a Zeeman field $m$ along the spin-polarized axis $\hat{z}$ and electrical potential $V$ (Appendix \ref{App_A1}): 
\begin{equation}
h_{0}(k_{x})=\hbar v_0(\sigma_{z}+t\sigma_{0})k_{x}+m \sigma_{z}-V\sigma_{0},  \label{ASMLSs}
\end{equation}
where $v_0$ is the Fermi velocity, $\sigma_z$ and $\sigma_0$ are the $z$-Pauli matrix and identity matrix in spin space. 
To induce asymmetry in the spin-momentum locking, the $tk_x\sigma_0$ term ($0<t<1$) tilts the dispersion [Fig. \ref{setup}(b)], resulting in distinct spin-dependent counter-propagating velocities, i.e. $v_{e}^{\sigma} = v_{0}(\sigma+t)$ where $\sigma = \pm 1$ denotes the two opposite spins. 

Systems described by Eq. (\ref{ASMLSs}) are suitable candidates for FEJD. 
The spin-momentum locking from the $k_{x}\sigma_{z}$ term breaks inversion symmetry while the asymmetry induced by $tk_x$ breaks TRS, which is usually achieved through the Zeeman field $m$.
The simultaneous breaking of both symmetries indicates the possibility of realizing the Josephson diode without a Zeeman field (Appendix \ref{App_A2}).

The center-of-mass momentum of paired electrons in ASMLS is crucial to achieve FEJDs.
This is modeled through the Bogoliubov-de Gennes (BdG) Hamiltonian \cite{Gennes1966} (Appendix \ref{App_A3})
\begin{equation}
H_{\text{BdG}}=\left( 
\begin{array}{cc}
h_{0}(k_{x}) & -i\Delta \sigma_{y} \\ 
i\Delta \sigma_{y} & -[h_{0}(-k_{x})]^{\ast }%
\end{array}%
\right) \text{,}  \label{HJJ}
\end{equation}%
in Nambu space, where $\Delta $ is the proximity-induced $s$-wave SC order parameter. 
A non-vanishing center-of-mass momentum of paired electrons \cite{Davydova2022}
\begin{equation}
k_{C}=\frac{1}{2}( k_{h}^{-\sigma}-k_{e}^{\sigma }) =\frac{1%
}{\hbar v_{0}}\frac{tV+m}{1-t^{2}}\text{,}  \label{kc}
\end{equation}
results from the net Fermi momentum between holes $k_{h}^{-\sigma}$ and electrons $k_{e}^{\sigma }$ [Fig. \ref{setup}(b)] with $k_{e( h) }^{\sigma }=+(-) (V-\sigma m)/[\hbar v_{0}(\sigma +t)]$ obtained by the electron and hole block dispersion in Eq. (\ref{HJJ}).

The momentum $k_C$ shifts the SC-gap center from $E=E_F=0$ to a finite energy, known as the Doppler energy shift \cite{Davydova2022} (Appendix \ref{App_A4}),
\begin{equation}
D = (1-t^{2})\frac{\hbar v_{0}k_{C}}{\Delta _{\text{eff}}} =\frac{tV+m}{\Delta _{\text{eff}}}
\text{,}  \label{DES}
\end{equation}
in the unit of effective SC gap $\Delta _{\text{eff}}=\sqrt{1-t^{2}}\Delta $.
As a result, the energy of right-propagating (left-propagating) spin-up (spin-down) states increases (decreases), manifesting as spin-dependent SC gaps $\Delta_{\sigma}=\Delta _{\text{eff}}( 1 - \sigma D)$ below $E_F=0$ [Fig. \ref{setup}(c)].

From Eq. (\ref{kc}) and (\ref{DES}), the joint effect of the asymmetry and gate voltage, $tV$, is seen to play the same role as the Zeeman field $m$, leading to gate-tunable center-of-mass momentum of paired electrons $k_C$.


\subsection{Transport properties of field-effect Josephson diode} \label{sec_kc_genfor}


Having explained how gate-tunable center-of-mass momentum of paired electrons $k_{\text{C}}$ can be achieved, we next describe how they give rise to FEJDs with gate-controlled diode efficiency $\eta$. 

Our prototypical FEJD setup, based on Eq. (\ref{HJJ}), is schematically illustrated in Fig. \ref{setup}(a) with super gates ($V_{\text{SC}}$) applied to the left ($x<0$) and right ($x>d$) SC leads and a tunneling gate ($V_{\text{TG}}$) applied to the middle non-SC region of length $d$ \cite{Mazur2022}.
The SC gap and electrical potential profiles are $\Delta(x)=\Delta\left[ e^{i\varphi _{L}}\Theta (-x)+e^{i\varphi _{R}}\Theta (x-d)\right]$ and $V(x)=V_{\text{SC}}\Theta (x^{2}-xd)+V_{\text{TG}}\Theta (xd-x^{2})$ with macroscopic SC phases $\varphi _{\text{L/R}}$ in the left/right SC leads. 

Here the supercurrent is carried by ASMLSs [Fig. \ref{setup}(a)]. 
Because of spin-momentum locking, the spin-up (spin-down) electrons are right-propagating (left-propagating), and undergo spin-flip Andreev reflections at the interfaces with the SC regions. 
Considering the scattering matrix at these interfaces (Appendix \ref{App_B}), the asymmetric BdG dispersion [Fig. \ref{setup}(c)] in the SC regions results in asymmetric Andreev reflection probabilities $R_{A,\pm}( E) \not=$\ $R_{A,\pm }( -E) $ for a quasiparticle of energy $E$, as shown in Fig. \ref{setup}(d), where 
\begin{equation}
R_{A,\pm  }(E)=|e^{-i\theta _{\pm  }}|^2 \text{,} \label{ra} 
\end{equation}
\begin{equation}
\theta _{\pm  }=\left\{ 
\begin{array}{ll}
\cos ^{-1}(\frac{E}{\Delta _{\text{eff}}}\mp D) & \text{, }%
\left\vert \frac{E}{\Delta _{\text{eff}}}\mp D\right\vert <1 \\ 
-i\cosh ^{-1}(\frac{E}{\Delta _{\text{eff}}}\mp D) & \text{, }%
\left\vert \frac{E}{\Delta _{\text{eff}}}\mp D\right\vert >1%
\end{array}%
\right. \text{.}
\end{equation}
%
%
In the $\left\vert \frac{E}{\Delta _{\text{eff}}}\mp D\right\vert <1$ case, $R_{A,\pm}=1$ identically.
This asymmetry crucially arises from the nonzero Doppler energy shift $D$ and gives rise to spin-dependent Andreev bound states (ABSs) spectra, 
which, in the short junction limit of $d\ll \hbar v_0/\Delta$, (Appendix \ref{App_B3}) is
\begin{equation}
E_{\pm }(\varphi )/\Delta _{\text{eff}}=\pm D\mp sgn[\sin \left( \varphi /2\right) ] \cos (\varphi
/2) \text{,}  \label{eabs1}
\end{equation}%
as plotted in Fig. \ref{setup}(e), where $\varphi =\varphi
_{R}-\varphi _{L}$ is the phase difference between two SC leads.
Importantly, gapless ABSs occurs when $\varphi=\varphi _{c}=2\cos ^{-1}D$.

Through the scattering matrix formalism \cite{Beenakker1991,Beenakker1992}, 
(Appendix \ref{App_C}), we obtain the zero-temperature supercurrent, 
\begin{equation}
I( \varphi ) =\left\{ 
\begin{array}{ll}
I_{\text{ABS}}( \varphi ) +I_{\text{cont}}  & \text{, } \vert
D\vert <1 \\ 
( e\Delta _{\text{eff}}/2\hbar) sign\left( D\right) \Gamma (
\varphi )  & \text{. } \vert D\vert \geq 1%
\end{array}%
\right. \text{,}  \label{i1}
\end{equation}%
where 
\begin{equation}
    I_{\text{ABS}}( \varphi ) =\frac{e\Delta _{\text{eff}}}{2\hbar }P(\varphi )\sin
(\varphi /2) \text{,} \label{iabs}
\end{equation}
with the ground-state parity $P(\varphi )=sign\{\cos (\varphi /2)-Dsign[\sin (\varphi /2)]\}$
, is the contribution from the ABS and 
\begin{equation}
    I_{\text{cont}} =\frac{e\Delta _{\text{eff}}}{2\hbar } \frac{2D}{\pi}  \text{,} \label{icont}
\end{equation}
arise from the continuum states. 
Particularly, when $m=0$, $I_{\text{cont}}= et V_{\text{SC}}/(\pi\hbar)$ is solely gate-induced.
For $\vert D\vert \geq 1$, both contributions become inseparable since $\varphi_c$ disappears, and the continuous expression is  $\Gamma (\varphi ) = (2/\pi) \left\{|D|-\sqrt{D^{2}-1}-\sin (\varphi /2)\tan ^{-1}\left[\frac{\sin (\varphi /2)}{\sqrt{D^{2}-1}}\right] \right\}$. 

As evident in Fig. \ref{setup}(f), $I(\varphi)$ exhibits a sharp discontinuity at $\varphi=\varphi_c$ for $D<1$. 
This has significant implications for the dissimilarity of the direction-selective critical Josephson currents, which are defined as the extrema $I(\varphi)$ i.e. $I_{c+}=\max_{\varphi }I(\varphi)$ and $I_{c-}=|\min_{\varphi }I(\varphi )|$.
In particular, for $0<D<1$, the current reverses at $\varphi=\varphi_c$ and $I_{c+}=I( \varphi_c )$, $I_{c-}=-I( \pi )$. 
Intuitively, we expect the ABSs contributions to $I_{c\pm}$ by the right-moving spin-up elections and the left-moving counterpart to be different because of the different gaps experienced, i.e. $I_{\text{ABS}}(\varphi_c)\propto \Delta_{+ }< |I_{\text{ABS}}(\pi)|\propto \Delta_{- }$.
Indeed, in the $D\ll 1$ limit, the $I_{\text{cont}}$ contribution is seen to enhance (suppress) the right-moving (left-moving) current, leading to $I_{c\pm }=e\Delta _{\text{eff}}/(2\hbar )(1\pm 2D/\pi )$. 

All in all, by substituting $I_{c\pm}$ into Eq. (\ref{etadef}), we obtain the diode efficiency
\begin{equation}
\eta =\frac{D}{|D|} \times \left\{ 
\begin{array}{ll}
\frac{4|D|/\pi +\sqrt{1-D^{2}}-1}{\sqrt{1-D^{2}}+1} & \text{, }|D|<1 \\ 
\frac{2(|D|-\sqrt{D^{2}-1})-\cot ^{-1}\sqrt{D^{2}-1}}{\cot ^{-1}\sqrt{D^{2}-1%
}} & \text{, }|D|>1%
\end{array}%
\right.  \label{eta1}
\end{equation}%
which solely depends on $D$ as shown in Fig. \ref{setup}(g), alongside results at nonzero temperatures $T$.
Notably, for $T=0$, the diode efficiency reaches a theoretical maximum of $|\eta |\sim 40.5\%$ at $|D|$ $=4\pi /(4+\pi ^{2})\approx 0.906$, which is comparable to the value obtained by applying a magnetic field \cite{Davydova2022, Lu2022}.
However, it gradually vanishes as $\left\vert D\right\vert $ gets larger, since in the $|D|\gg 1$ limit, the extrema of the current $\lim_{\left\vert D\right\vert \gg 1}I(\varphi)=e\Delta _{\text{eff}}/(hD)\cos \varphi $ becomes symmetric, as shown in Fig. \ref{setup}(f). 

The exclusive functional dependence of the diode efficiency $\eta$ on the Doppler energy shift $D=(tV_{\text{SC}}+m)/\Delta_{\text{eff}}$ [Eq. (\ref{DES})] implies that the gate control voltage $V_{\text{SC}}$ plays, in the presence of asymmetry $t$, an indistinguishable role from that of the Zeeman field $m$ \cite{Davydova2022}. 
Hence, the diode efficiency can be completely gate-controlled; for instance, in the $m=0$ and $D\ll 1$ limit, $\eta \propto D \propto tV_{\text{SC}}$ (Appendix \ref{App_C2}). 
\begin{figure}
\centering \includegraphics[width=0.4\textwidth]{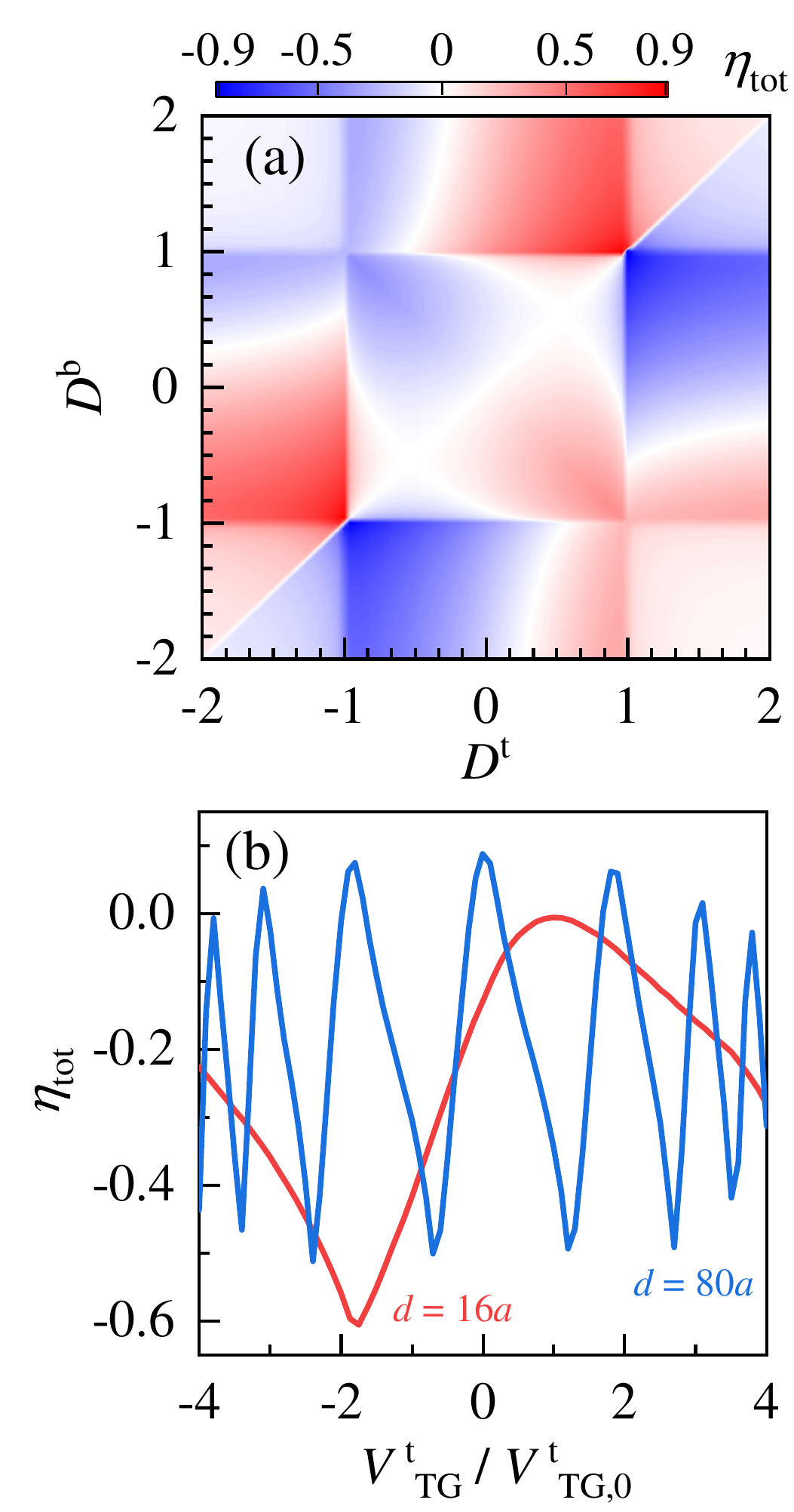}
\caption{
Diode efficiency of a topological FEJD through two-edge interference in a TRS-broken QSHI. 
(a) The interfering diode efficiency maximized to $|0.9|$ when $|D^{\text{t/b}}|\gtrsim 1$ and $|D^{\text{b}}|\lesssim 1$.
Other parameters are the same as those in Fig. \ref{setup}. 
(b) The $V^{\text{t}}_{\text{TG}}$-controlled oscillating diode efficiency originates from the phase difference between two currents. 
$a=1$ is the lattice constant and $V^{\text{b}}_{\text{TG}}=0$.
}
\label{topjd}
\end{figure}

\section{Topological Feild-effect Josephson diode} \label{sec_topfejd}
The proposed FEJD might be realized in a specific type of QSHI with \textit{asymmetric} helical edge states serving as the ASMLS in FEJDs.
The possibility of realizing asymmetric helical edge states in QSHI is demonstrated in Sec. \ref{sec_topfejd_asmls}.
The helical edge states are protected by a spin-up-down parity symmetry $\hat{S}$, rather than the time-reversal symmetry \cite{Sheng2005, Li2012, Li2013}. 
As a result, when the spin-Chern number of the system is well-defined and non-zero, the asymmetric helical edge states \cite{Fu2022} are expected in Floquet topological insulator \cite{Fu2022} or magnetic-doped QSHI \cite{ Chen2021, Zhang2019}.

QSHIs are ideal superconducting quantum interference devices that are proposed as a Josephson diode  \cite{Souto2022,Fominov2022}.
By combining the effect of supercurrent interference and ASMLSs, the diode efficiency is greatly enhanced to $90\%$, which is discussed in Sec. \ref{sec_topfejd_scint}.
When considering a finite-length junction, the effect of the tunneling gate ($V_{\text{TG}}$) becomes measurable, which affects the diode efficiency by tuning the interference current as illustrated in Sec. \ref{sec_topfejd_vtg}.


\begin{figure*}[t]
\centering \includegraphics[width=0.85\textwidth]{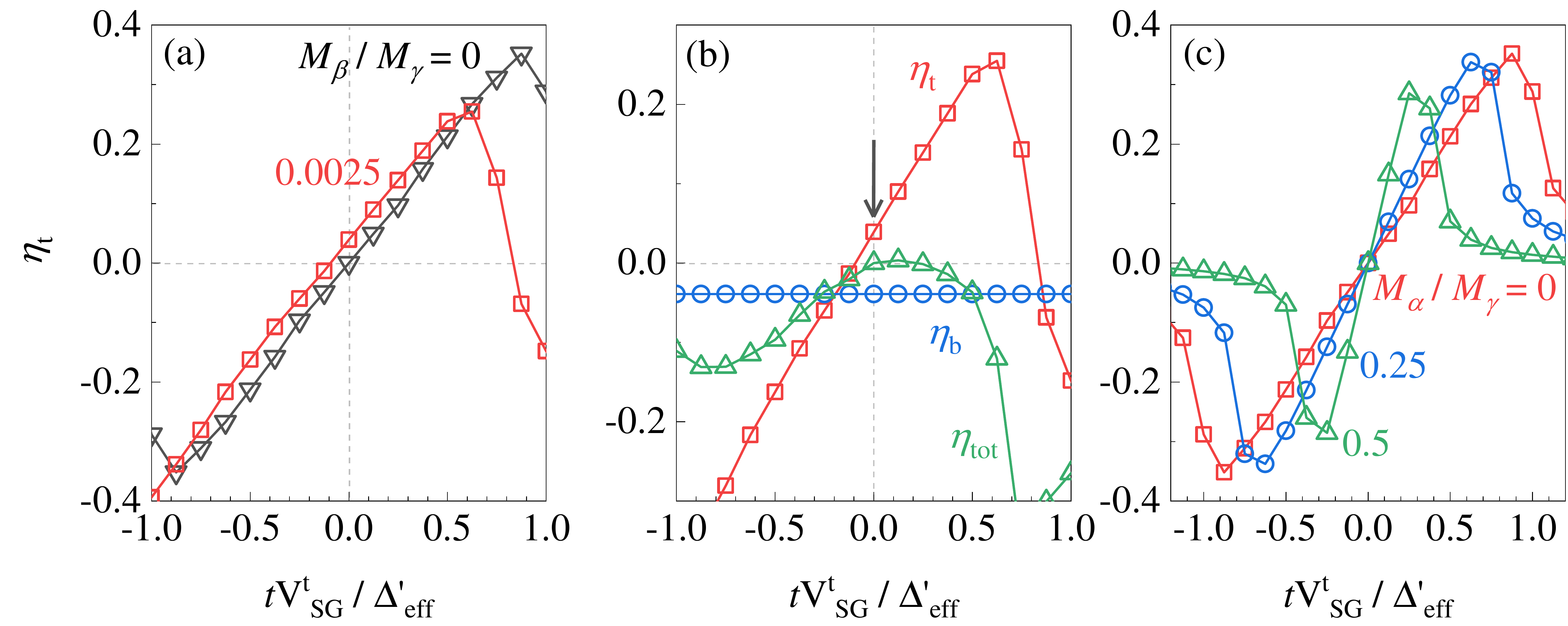}
\caption{The effect of $M_{\beta }$ on (a) single edge, and (b) double edge.
Here we set, $\Delta ^\prime_{\text{eff}}=2\sqrt{1-t^{2}}\Delta(T)$. 
(c) The effect of $M_{\alpha}$ on a single-edge Josephson diode (top edge).
An increasing $M_{\alpha}$ reduces the edge-state SC gap resulting in a lower gate voltage ($tV^{\text{t}}_{\text{SV}}/\Delta^\prime _{\text{eff}}\sim1$) to reach the maximum diode efficiency. 
}
\label{mb}
\end{figure*}

\subsection{Asymmetric helical edge states in quantum spin Hall insulator} \label{sec_topfejd_asmls}
Before diving into the field-effect behavior of the Josephson diode, we first illustrate the existence of asymmetric helical edge states in QSHI without TRS. A general ansatz of QSHI Hamiltonian is \cite{Fu2022,Chen2021,Zhang2019,Bernevig2006,Chen2021,Zhang2019}
\begin{eqnarray}
H_{\text{QSHI}}(\bm{k}) &=&\left( 
\begin{array}{cccc}
M_{\gamma } & 0 & 0 & \hbar v_0k_{-} \\ 
0 & -M_{\gamma } & \hbar v_0k_{-} & 0 \\ 
0 & \hbar v_0k_{+} & M_{\gamma } & 0 \\ 
\hbar v_0k_{+} & 0 & 0 & -M_{\gamma }%
\end{array}%
\right)  \label{hqshi} \\ 
&+&\left( 
\begin{array}{cccc}
M_{+} & 0 & 0 & t\hbar v_0k_{-} \\ 
0 & M_{-} & -t\hbar v_0k_{-} & 0 \\ 
0 & -t\hbar v_0k_{+} & -M_{+} & 0 \\ 
t\hbar v_0k_{+} & 0 & 0 & -M_{-}%
\end{array}%
\right) \text{,} \nonumber 
\end{eqnarray}
where $\bm{k}=(k_{x},k_{y})$, $k_{\pm }=k_{x}\pm ik_{y}$, $M_{\gamma }=M_{\gamma
}-B_{\gamma }k_{+}k_{-}$, and $M_{\pm }=M_{\alpha }\pm M_{\beta }$.
This Hamiltonian is written in the basis $\{\vert s,\uparrow \rangle ,\vert p,\uparrow \rangle,\vert s,\downarrow \rangle ,\vert p,\downarrow\rangle \} ^{T}$, where $s$ ($p$) denotes the sublattice/orbital component and $\uparrow $ ($\downarrow $) denotes the spin component of the bulk states which are distinct from the spin alignments of the edge states.
The first term of Eq. (\ref{hqshi}) is the Bernevig-Hughes-Zhang Hamiltonian \cite{Bernevig2006} of a QSHI with TRS which is broken by the second term induced by high-frequency circularly polarized light \cite{Fu2022} or magnetic doping \cite{ Chen2021, Zhang2019}.
 
It is possible to maintain the gapless edge states in a TRS-broken QSHI if the system is invariant under the up-down parity symmetry $\hat{S}=\sigma_z\tau_z$, i.e. $[ \hat{S},H_{\text{QSHI}}] =0$.
With this symmetry, the Hamiltonian Eq. (\ref{hqshi}) transferred into a block-diagonalized form corresponding to bispinors with only two upper or lower nonzero components that describe a quantum anomalous Hall insulator whose topology is determined by the Chern number of each block as $C_{\pm }=\pm [ sign( M_{\gamma }+M_{\alpha })+sign( B_{\gamma }) ]/2$.
The spin-Chern number of the TRS-broken QSHI is thus well-defined as $C_{s}=(C_+-C_-)/2$.
The gapless edge perseveres even when TRS is broken.
With a non-zero $C_{s}$, the asymmetric helical edge states \cite{Fu2022} is described by 
\begin{equation}
h_{e}^{\text{s},\sigma }(k_{x})=s(\sigma \hbar v_{0}+\hbar
v_{t})k_{x}-V_{\text{SG}}^{\text{s}}+m_{\text{s}} \text{,} \label{edge} 
\end{equation}
where $\sigma$ represents the spin orientation of the edge state, s$=$t$\equiv +1$ (s$=$b$\equiv -1$) represents the top (bottom)-edge, and $V_{\text{SG}}^{\text{s}}$ is the SG applied locally on each edge.

\subsection{Supercurrent interference between two edges}\label{sec_topfejd_scint}

By specifying $h_{0}(k_{x})$ in Eq. (\ref{HJJ}) as Eq. (\ref{edge}), it is expected that the top (t) and bottom (b) edges of a QSHI each constitute an individual FEJD, with gates $V_{\text{SC}}^{\text{s}=\text{t/b}}$ and Zeeman fields $m^{\text{s}}=\pm m$ \cite{Fu2022}.
It is straightforward to apply Eqs. (\ref{i1}-\ref{eta1}) to calculate each edge current $I^{\text{s}}(\varphi )$ and the diode efficiency $\eta ^{\text{s}}$ (Appendix \ref{App_D2}).

A QSHI is an ideal SC quantum interference device where the supercurrent through a device results from the interference between two ASMLS currents.
As a result, the advantage of this topological implementation is to enhance $\eta$ through interference \cite{Souto2022} between two topological edge currents.
The combined efficiency is (Appendix \ref{App_D3}) 
\begin{equation}
\eta ^{\text{tot}}=\left\{ 
\begin{array}{ll}
\frac{\sqrt{1-(D^{\text{t}})^{2}}-\sqrt{1-(D^{\text{b}})^{2}}+\frac{2}{\pi}(D^{\text{t}%
}-D^{\text{b}}) }{\sqrt{1-(D^{\text{t}})^{2}}+\sqrt{1-(D^{\text{b}})^{2}}%
} & \text{, }0\leq D^{\text{t}}\leq1 \\ 
\frac{Z_{-}+2( D^{\text{t}}-D^{\text{b}}-\sqrt{(D^{\text{t}%
})^{2}-1})}{Z_{+}}    & \text{, }D^{\text{t}}>1%
\end{array}%
\right. \text{, } \label{etatot}
\end{equation}
with $0\leq D^{\text{b}}<1$ and $Z_{\pm }=\frac{\pi }{2}-\cot ^{-1}(\sqrt{(D^{\text{t}})^{2}-1})\pm \sqrt{%
1-(D^{\text{b}})^{2}}[ \frac{\pi }{2}+\tan ^{-1}( \sqrt{\frac{%
1-(D^{\text{b}})^{2}}{(D^{\text{t}})^{2}-1}}) ] $
shows that when $|D^{\text{t}}|\gtrsim1$ and $|D^{\text{b}}|\lesssim1$, the diode efficiency is enhanced up to $90\%$, much higher than the one in individual edge [Fig. \ref{topjd}(a)].
Notably, this enhancement is unattainable through a magnetic field \cite{Tkachov2015,Scharf2021,Dolcini2015}, which cancels \cite{Dolcini2015} continuum-state currents with $D^{\text{t}}=D^{\text{b}}$
, thereby leading to a vanishing diode effect.

\subsection{Finite-length junction and numerical simulations} \label{sec_topfejd_vtg}

In a finite-length junction, $d>\hbar v_0\Delta$, a phase difference between two interfering ASMLS is accumulated \cite{Fu2022}, which gives rise to $I^{\text{tot}}=\sum_{\text{s}}I^{\text{s}}(\varphi +\varphi _{\text{TG}}^{\text{s}})$ with $\varphi _{\text{TG}}^{\text{s}}=2tV_{\text{TG}}^{s}d[\hbar v_{F}(1-t^{2})]^{-1}$ and $\varphi _{\text{TG}}^{\text{t}}(V^{\text{t}}_{\text{TG},0})=2\pi$. 
As a result, the phase difference between two interfering currents is controlled by the TGs and the critical current is the consequence of constructive or destructive interference leading to a TG-controlled oscillating diode efficiency [Fig. \ref{topjd}(b)].

The result in a finite-length junction is numerically obtained through the lattice Green's function technique \cite{Xu2018,Fu2019,Fu2022}. 
We set
$M_{\gamma }=2B_{\gamma }$ ($B_{\gamma }=1$). 
The width of the junction is $W=30a$ ($a\equiv 1$ is the lattice constant) which is much larger than the characteristic lengths of the edge states ($\sim 5a$), to avoid the interactions between opposite edges. 
In the TRS-broken term, $M_{\alpha}=0.1M_{\gamma }$ to keep the system supporting ASMLSs with $t=0.1$ while $M_{\beta}=0$ is set for simplification.
In the SC regions, the SC gap is $\Delta (T)=\Delta (0)\tanh (1.74%
\sqrt{T_{c}/T-1})$, where $T_{c}$ is the critical temperature and $\Delta
(0)=0.01M_{\gamma }$ is the zero-temperature SC gap. 
The N-SC interfaces are assumed to be transparent. SGs and TGs are applied within the areas of a width comparable to the characteristic lengths of the edge states (Appendix \ref{App_D4}). 
When manipulating the gates, the Fermi levels are restricted within the bulk gap to ensure that the supercurrent is solely contributed by the edge states. 
A unit gate is defined
as $V_{TG,0}^{\text{s}}=\Delta (T)(1-t^{2})/t$ corresponding to the phase $\varphi
_{g0}^{\text{s}}=2tV_{TG,0}^{\text{s}}d_{0}[\hbar v_{F}(1-t^{2})]^{-1}=\pi $ with $%
d_{0}\approx 80a$. 
Only the cases with the gate at the top edge are studied while the gate at the bottom edge is turned off ($V_{\text{TG}}^{\text{b}}=0 $).

Apart from the asymmetry in helical edge states, the TRS-broken term in Eq. (\ref{hqshi}) also includes (i) the $M_{\beta }$ term, which is not related to the band topology but induces a spin-dependent shifting to the energy spectrum like a Zeeman
field \cite{Zou2012}, and (ii) the $M_{\alpha }$ term, which modifies the bulk band gaps and can finally, drive the system to be in a QAHI phase when $\vert M_{\alpha
}\vert >M_{\gamma }$. Since only the case with helical edge state is concerned here, the effect of the QAHI phase is beyond our discussion, and $\vert M_{\alpha
}\vert <M_{\gamma }$ is set. 

Our numerical results exhibit that effect (i) leads to a gate-free diode effect on each edge [Fig. \ref{mb}(a)], while in the whole system, the diode effect vanishes because of the compensation between two edges [Fig. \ref{mb}(b)]. 
The reason is the Doppler energy shift depends on $M_{\beta }$ via $m_{\text{s}}=$s$M_{\beta }$. As a result, when $V_{\text{SC}}^{\text{s}}=0$, we obtain $D^{\text{t}}=D^{\text{b}}$, and thus $\eta ^{\text{t}}=-\eta ^{\text{b}}$ and $\eta ^{\text{tot}}=0$. 
Effect (ii) is dual. 
First, the Josephson diode vanishes when the system is driven to be a QAHI, which is beyond this work due to the destruction of helical edge states. 
Besides, we observe that edge-state SC gap $\Delta _{\text{edge}}$ depends on both the bulk-state SC gap $ \Delta _{\text{bulk}}=\Delta(T)$ as expected and bulk-state band gap controlled by $M_{\alpha }$ which is reported in Ref. \cite{Tkachov2019a}. 
Since an increasing $M_{\alpha }$ reduces $\Delta _{\text{edge}}$, a lower gate voltage is required to obtain the maximum value of the diode efficiency on each edge as shown in Fig. \ref{mb}(c).

It is possible to induce a gap in the edge spectrum when the TRS is broken. This couples the left- and right-mover in the edge leading to a back-scattering that smoothens the skewed CPR and thus reduces the diode efficiency in each edge. 
However, because of the interference, the total diode efficiency can be enhanced (Appendix \ref{App_D4}).

\begin{figure*}
    \centering \includegraphics[width=0.85\textwidth]{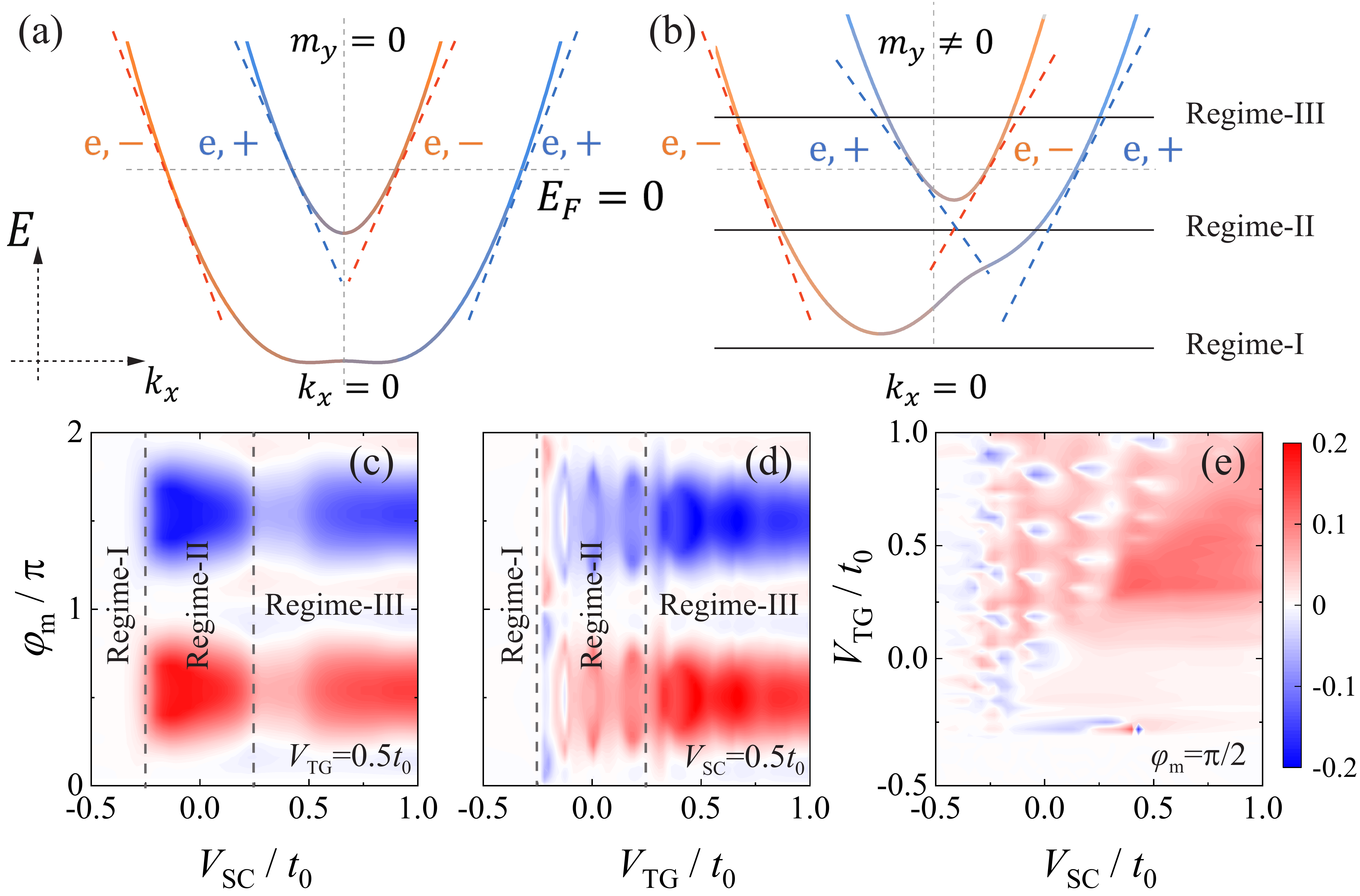}
    \caption{
    Mechanism and gate-tunable operation of semiconductor-based FEJD. 
    (a) and (b) show dispersions ($k_y=0$) of 2D semiconductors with pure RSOI $\alpha_R=t_0$ ($t_{0}=1$), $\alpha_D=0$ and Zeeman field $\bm{m}=m_{0}( \cos \varphi _{m},\sin \varphi_{m},0) $ with $m_0=0.1t_0$.
    ASMLSs occur around $E_F=0$ (the dashed grey horizontal lines) when $m_y=m_0\sin \varphi_{m}\neq0$. 
    The diode efficiency in (c-e) is attributed to the ASMLSs in three transport regimes experienced by the systems when sweeping the gates.  
    The parameters are $\Delta(0)=0.05t_0$, $T/T_c = 0.1$, and $d=20a$.
    }
    \label{phimVSGVTG}
\end{figure*}

\section{Field-effect Josephson diodes in semiconductors}\label{sec_semifejd}
The mechanism of realizing a FEJD using ASMLSs is also attainable in semiconductors. The ASMLSs occur in semiconductors due to the interplay between the Zeeman field and the SOIs. The general Hamiltonian in the basis $\{\psi_\uparrow,\psi_\downarrow\}^T$ is \cite{Rasmussen2016,Liu2010} 
\begin{equation}
    H_{\text{2D}}(\bm{k})=t_{0}a^{2} |\bm{k}|^2-V + h_{\text{so}}
     +\bm{m}\cdot \bm{\sigma} \text{,} \label{h2deg}
\end{equation}
where $h_{\text{so}}=\alpha _{\text{R}}a (\bm{\sigma}\times \bm{k})_z+\alpha _{\text{D}}a( k_{x}\sigma _{x}-k_{y}\sigma _{y}) $ describes the Rashba-SOI (RSOI) $\alpha _{\text{R}}=\alpha _{0}\sin \varphi _{\text{so}}$ and Dresselhaus-SOI (DSOI) $\alpha _{\text{D}}=\alpha _{0}\cos \varphi _{\text{so}}$ with $\varphi_{\text{so}}$ denoting the proportion of these two components. $\bm{m}=m_{0}( \cos \varphi _{m},\sin \varphi_{m},0) $ is the in-plane Zeeman field. 

Two pairs of ASMLSs occur near $E_{F}$ [Fig. \ref{phimVSGVTG}(a) and (b)] (Appendix \ref{App_E2}), resembling two edges in TRS-broken QSHIs.
This is because, for example, 
when $\alpha _{\text{D}}=0$ and $\varphi_{m}\neq n\pi$ ($n\in\mathbb{Z}$), dispersion of $H_{\text{2D}}(\bm{k})$ is shifted in $k_x$ direction. 
For the cases with other $\alpha _{\text{D}}$ or $\varphi_{so}$, the shift of dispersion in momentum space is induced by different $\varphi_{m}$.
Thus, a gate-tunable center-of-mass momentum of paired electrons is expected in a semiconductor-based FEJD. 

Diode efficiencies of FEJD in the $x$-direction exhibit three operating regimes (Appendix \ref{App_E3}) depending on the gate-controlled Fermi level [Fig. \ref{phimVSGVTG}(c-e)]. 
In Regimes I/II/III, none/one/two ballistic SC channels across the Fermi level [Fig. \ref{phimVSGVTG}(b)]. 
$\eta $ is absent in Regime I and oscillates with $V_{\text{TG}}$ in Regime II due to the Fabry-P\'{e}rot interference between one ballistic and one tunneling channel [Fig. \ref{phimVSGVTG}(d)] resembling the topological implementation in Fig. \ref{topjd}(b).

Although the mechanism of FEJD is shared with the topological FEJD due to the ASMLSs, the gate-operating performance exhibited in Fig. \ref{phimVSGVTG}(c-e) is not as good as the topological implementation, because of the inevitable back-scattering between $|\text{e},\pm\rangle$ and $|\text{e},\mp \rangle$ due to the parallel spin-orientation component, which is forbidden in QSHI where spin orientation is tightly locked to the momentum \cite{Yokoyama2013,Yokoyama2014}.

The formation of ASMLSs also accounts for the dependence of the diode efficiency on the direction of the in-plane Zeeman field. 
With pure RSOC ($\varphi_{so}=\pi/2$), the diode efficiency vanishes when $\varphi_{m}= n\pi$, because of the symmetric dispersion [Fig. \ref{phimVSGVTG}(a)].
But when the in-plane Zeeman field is parallel to $k_{x}\sigma_y$, diode efficiency is maximized since the center-of-mass momentum of paired electrons is pronounced due to the maximum shift of the dispersion.

Although our results are qualitatively consistent with the experiment observations \cite{Mazur2022}, there are some deviations in the $\eta -\varphi _{m}$ relation which only indicates the direction of the pure SOI field and is not a simple sinusoidal function \cite{Mazur2022}, but contains higher-harmonic terms (Appendix \ref{App_E4}).

\section{Discussions and conclusions} \label{sec_discon}

As for experimental realization, semiconductor-based Josephson diodes are widely investigated in InAs \cite{Baumgartner2022a,Baumgartner2022b,Costa2022} and InSb \cite{Turini2022,Mazur2022} nanostructures.
In these systems, ASMLSs are expected because of the interaction between the SOI and the in-plane magnetic field and FEJD is expected as we illustrated above.
In particular, in a proximity-magnetized Pt layer by ferrimagnetic insulating Y$_3$Fe$_5$O$_{12}$, where the ferromagnetic proximity achieves the Zeeman effect \cite{Jeon2022}, a \textit{purely} free-field FEJD is expected by some modifications in the current experiment setup.

Candidates of the topological FEJD are magnetic-doped QSHIs such as HgTe \cite{Hart2014,Liu2008}, MnBi$_{2}$Te$_{4}$ \cite{Chen2022,WZXu2022}, and WTe$_{2}$ \cite{Huang2020,Lupke2020}.
The asymmetric helical edge state may be induced by a small fixed magnetic field polarizing the magnetic atoms along a specific crystallographic $z$-direction.
The typical SC proximate gaps are in the order from $0.1$ meV in HgTe \cite{Hart2017,Bocquillon2017} and MnBi$_{2}$Te$_{4}$ \cite{Chen2022,WZXu2022} to $1$ meV in WTe$_{2}$ \cite{Huang2020,Lupke2020}.
To achieve the optimized diode efficiency with $D\approx0.91$, electrical potential around $10$ meV \cite{Simoni2019,Xu2021} is required for $t=0.1$, which guarantees the edge-state supercurrent by pining Fermi level within band gap ($\sim30$ meV).
Corresponding critical currents ($\sim 10 $ nA) are experimentally detectable \cite{Baumgartner2022a,Baumgartner2022b}.
Although the existence of asymmetric helical edge states is under exploration, we believe a topological FEJD could be realized shortly thanks to the development of material science.

From the two implementations of QSHI and semiconductor, we expect that the mechanism of FEJD induced by ASMLSs is universal. 
Although only the one-dimensional ASMLSs are studied here, similar mechanisms are also valid in systems supporting 2D/3D tilted Direc/Weyl cones such as the topological surface states in a 3D topological insulator \cite{Zheng2020,Zhang2021} and bulk states of type-I and type-II Dirac/Weyl semimetals \cite{Pal2022}. 
The occurrence of ASMLSs in semiconducting systems is not restricted to the one with DSOC and RSOC that are linearly dependent on the wave vector. 
It is also possible to realize FEJD in more advanced systems such as semiconductors with cubic spin-orbit coupling \cite{Alidoust2021}.

The mechanism and the proposed FEJD are valid even if the effect of charge redistribution \cite{Chen2012,Bahramy2012,Stehno2020} is taken into account.
The charge redistribution occurs at the interface between the SC leads and the central non-SC section when the Fermi levels in these regions are mismatched because of different $V_{\text{TG}}$ and $V_{\text{SG}}$.
With the charge redistributed, the resulting bend-bending and even band inversion effects are possible to obstacle the supercurrent non-reciprocity.
The realistic gate-controlled behaviors of EFJD involving these effects are beyond this work and deserve further studies based on the solutions of the self-consistent Schr\"odinger–Poisson equations \cite{Stehno2020}.
However, the main results of the current work are valid. 
The reasons are as follows.
(i) FEJD can be achieved by solely controlling $V_{\text{SG}}$ and keeping $V_{\text{TG}}=V_{\text{SG}}$. 
As a result, no charge redistribution occurs due to the identical Fermi level between the SC and non-SC regions.
In particular, in the short-junction limit, the effect of TGs in modulating the anomalous Josephson phase is negligible.
(ii) In the topological FEJDs, the supercurrent carried by the asymmetric helical edge states is robust against non-magnetic potential at the interface. The effect of the charge redistribution is thus negligible, although it might be pronounced when the bulk states are involved \cite{Stehno2020}. 
(iii) In the semiconductor-based FEJD, the effect of the charge redistribution may become measurable and reduce the diode efficiency, but the gate-controlled behaviors of EFJD still exist \cite{Mazur2022}.

In conclusion, we proposed a new mechanism to achieve FEJD, which originates from the gate-tunable center-of-mass momentum of paired electrons in ASMLS.
The mechanism is schematically illustrated in Fig. \ref{setup}.
Based on this mechanism, we proposed a topological FEJD in a specific type of TRS-broken with asymmetric helical edge states. The diode efficiency is enhanced to $90\%$ due to the supercurrent interference effect between two edges. To the best of our knowledge, this is the highest theoretically predicted value in two-terminal Josephson junctions. 
We further demonstrate that The mechanism of realizing FEJDs via ASMLSs is also attainable in semiconductors subjected to the interplay between the Zeeman field and the SOI. 
Our mechanism results provide an alternative explanation to the recent experiment \cite{Mazur2022} and we expected that FEJD could be realized by some modifications in the current experiment setup.
Based on these aspects, we believe our mechanism and FEJD can be achieved in future experiments and have potential applications in topological superconducting computation and cryogenic electronics.

\begin{acknowledgments}

P.-H. F. thanks Weiping Xu, Shi-Han Zheng, Shuai Li, Jiayu Li, and Jin-Xin Hu for inspiring discussions. 
P.-H. F. \& Y. S. A. are supported by the Singapore Ministry of Education (MOE) Academic Research Fund (AcRF) Tier 2 Grant (MOE-T2EP50221-0019). 
C.H.L. acknowledges support from Singapore’s NRF Quantum engineering grant NRF2021-QEP2-02-P09 and Singapore’s MOE Tier-II grant award No: MOE-T2EP50222-0003.
J.-F. L. is supported by the National Natural Science Foundation of China (Grants No. 12174077), the Bureau of Education of Guangzhou Municipality (Grant No. 202255464), and the Natural Science Foundation of Guangdong Province (Grant No. 2021A1515012363).

\end{acknowledgments}

\appendix
\section{Features of the asymmetric spin-momentum-locked Hamiltonian} \label{App_A}
In this appendix, we demonstrate the properties of the ASMLSs Hamiltonian including (i) the asymmetric spin-momentum-locking natures revealed in the group velocity of the electrons, 
(ii) the general requirements for the Josephson diode effect satisfied by the Hamiltonian, 
(iii) the gate-tunable a non-zero center-of-mass velocity and 
(iv) Doppler energy shift.

\subsection{Hamiltonian, dispersion and asymmetric\ group velocity} \label{App_A1}

\begin{figure}[t]
\centering \includegraphics[width=0.5\textwidth]{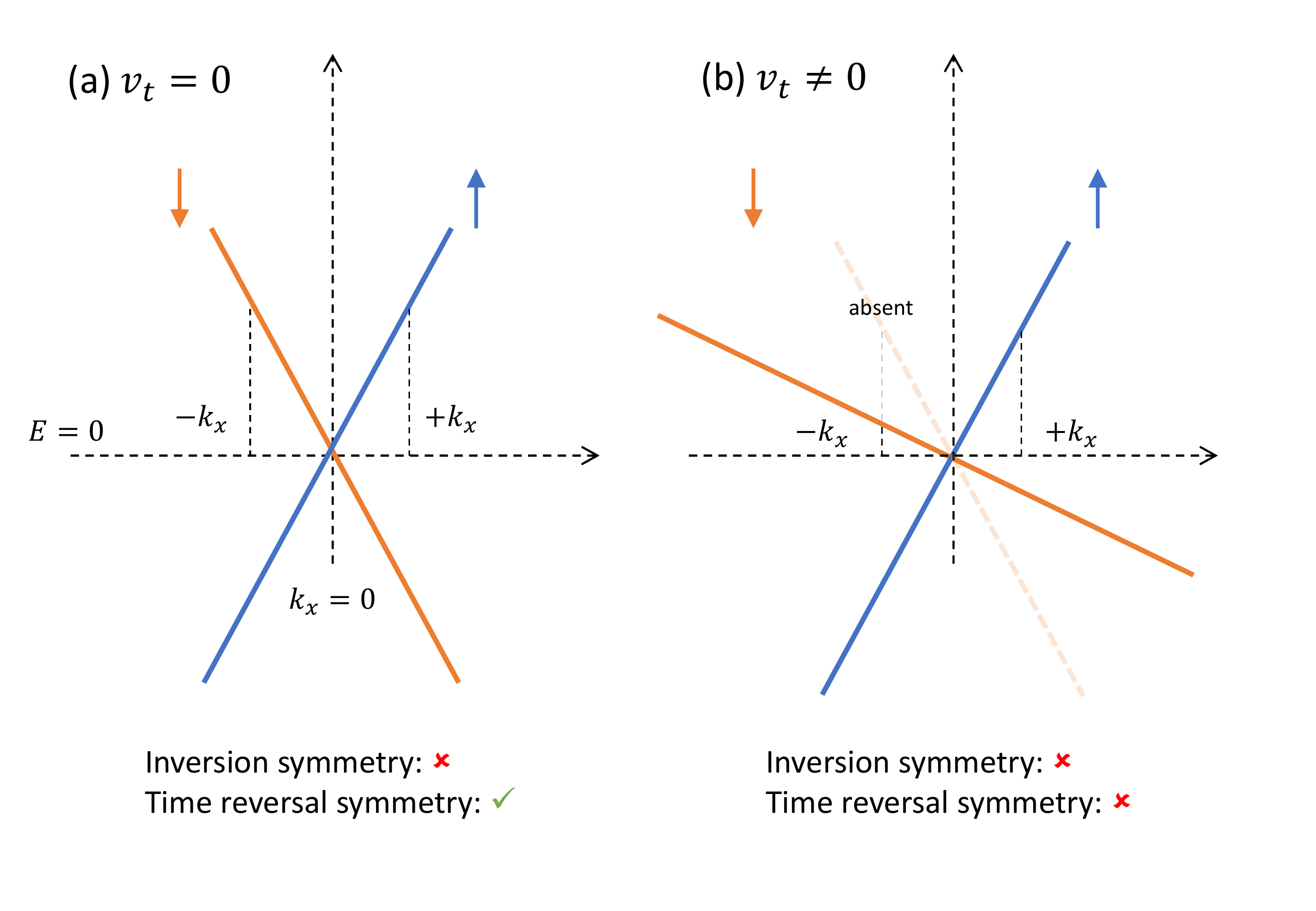}
\caption{Schematic dispersion of Eq. (\ref{asmldis}) with (a) $v_t=tv_0=0$ and (b) $v_t\neq0$. The spin-momentum locking nature requires a breaking of inversion symmetry and the asymmetry breaks the time-reversal symmetry.
}
\label{rdis}
\end{figure}

As exhibited in Fig. \ref{rdis}, the dispersion of Eq. (\ref{ASMLSs}), 
\begin{equation}
E_{e}^{\sigma }\left( k_{x}\right) =\hbar v_{0}\left( \sigma +t\right)
k_{x}-V+\sigma m\text{,}  \label{asmldis}
\end{equation}%
is tilted by $t$, resulting in an asymmetric group velocity of the electrons%
\begin{equation}
v_{e}^{\sigma }=\frac{\partial E_{e}^{\sigma }}{\partial k_{x}}=v_{0}\left(
\sigma +t\right) \text{,}
\end{equation}%
where the $\sigma v_{0}$ term indicates the spin-momentum-locking feature, namely, electrons propagating in spin-selective directions, and the $tv_{0}=v_t$ term indicates the asymmetry -- electrons with different spins
counter-propagating with distinct velocities.

\subsection{General requirements for Josephson diode effect} \label{App_A2}

The breaking of time-reversal and inversion symmetries is the general requirement for the observation of the Josephson diode effect. The dispersion Eq. (\ref{asmldis}) with $m=0$ is shown in Fig. \ref{rdis}. 
When $v_{t}=0$ [Fig. \ref{rdis}(a)], the inversion symmetry $\mathcal{I}$ is broken because states transform from $+k_{x}$ to $-k_{x}$ without a spin-flipping is prohibited, but the time-reversal symmetry is $\mathcal{T}$ is preserved because it is allowed to transform from $+k_{x}$ to $-k_{x}$ with a spin-flipping $+\sigma $ to $-\sigma $. 
When $t\neq 0$ [Fig. \ref{rdis}(b)], both $\mathcal{I}$ and $\mathcal{T}$ are broken because when flipping the spin from $+\sigma $ to $-\sigma $ the $-k_{x}$ state is absent because of $tv_{0}$.

Therefore, the requirement for the observation of the Josephson diode effect is met in our proposal because (i) the spin-momentum-locking nature requires a breaking of inversion symmetry, and (ii) the asymmetry breaks the time-reversal symmetry. Thus, the Josephson diode effect is expected in an ASMLS. 

\subsection{The non-zero center-of-mass momentum} \label{App_A3}

When considering the superconducting proximity effect, the one-dimensional BdG Hamiltonian [Eq. (\ref{HJJ})] can be written in Nambu space $(\psi _{\sigma},\psi _{\bar{\sigma}}^{\dagger })^{T}$ as
\begin{equation}
\hat{H}_{BdG}^{\sigma }(k_{x})=\left( 
\begin{array}{cc}
h_{e}^{\sigma }\left( k_{x}\right) & \sigma \Delta \\ 
\sigma \Delta ^{\dagger } & h_{h}^{\bar{\sigma}}\left( k_{x}\right)%
\end{array}%
\right) \text{,}  \label{1dbdg}
\end{equation}%
where 
\begin{eqnarray}
h_{h}^{\bar{\sigma}}\left( k_{x}\right) &=&-\left[ h_{e}^{\bar{\sigma}}\left(
-k_{x}\right) \right] ^{\ast }\nonumber \\
&=&\hbar v_{0}\left( -\sigma +t\right)
k_{x}+\sigma m+V\text{,}  \label{hh}
\end{eqnarray}%
is the Hamiltonian of the paired spin-$\bar{\sigma}$ holes with $\bar{\sigma}=-\sigma $. 
From Eq. (\ref{asmldis}) and (\ref{hh}), the Fermi momentum of the electrons and holes are determined by $E_{e}^{\sigma }\left( k_{e}^{\sigma}\right) =E_{h}^{\sigma }\left( k_{h}^{\sigma }\right) =E_{F}=0$, which are 
\begin{equation}
k_{e\left( h\right) }^{\sigma }=+\left( -\right) \frac{V-\sigma m}{\hbar
v_{0}\left( \sigma +t\right) }\text{.}
\end{equation}%
The non-zero center-of-mass momentum is defined by the net Fermi momentum between the
paired electrons and holes as shown in Eq. (\ref{kc}).

\subsection{The Doppler energy shift} \label{App_A4}

The dispersion of Eq. (\ref{1dbdg}) is 
\begin{eqnarray}
    E_{\sigma ,\pm }\left( k_{x}\right) &=&\left( t\hbar v_{0}k_{x}+\sigma
m\right) \nonumber \\
&&\pm \sqrt{(\sigma \hbar v_{0}k_{x}-V)^{2}+\Delta ^{2}}\text{,} \label{disp}
\end{eqnarray}
which is shown in Fig. \ref{setup}(c) with the band edges of the upper($+$)/lower($-$) spin-$\sigma $ are 
\begin{equation}
\Delta _{\sigma ,\pm }=E_{\sigma ,\pm }\left( k_{\sigma ,\pm }\right)
=\sigma \left( tV+m\right) \pm \sqrt{1-t^{2}}\Delta \text{,}
\end{equation}%
where 
\begin{equation}
k_{\sigma ,\pm }=\sigma \frac{1}{\hbar v_{0}}\left( V\mp \frac{t\Delta }{%
\sqrt{1-t^{2}}}\right)
\end{equation}
determines the extreme value of the dispersion curve satisfying $\partial_{k_{x}} E_{\sigma ,\pm }( k_{\sigma ,\pm }) =0$. Two modifications are induced by the asymmetry:

(i) The superconducting band gap. Defined by the difference between the upper and lower band, the superconducting band gap is 
\begin{equation}
\left\vert \Delta _{\sigma ,+}-\Delta _{\sigma ,-}\right\vert =2\sqrt{1-t^{2}%
}\Delta =2\Delta _{\text{eff}}\text{,}
\end{equation}%
which is modified from $2\Delta $ to $2\Delta _{\text{eff}}$. T
he effective band gap $\Delta _{\text{eff}}$\ is defined for simplification.

(ii) The Doppler energy shift. The band gap center is shifted from $E_{F}=0$ to 
\begin{eqnarray}
    \epsilon _{\sigma ,+}\left( t,V,m\right) -\epsilon _{\sigma ,+}\left(
0,0,0\right) &=&\sigma \left( tV+m\right) \\
&=&\sigma \left( 1-t^{2}\right)
\hbar v_{0}k_{C} \nonumber\text{,} \label{bandedge}
\end{eqnarray}
which is known as the so-called Doppler energy shift induced by $k_C$. 
Rewritten in the unit of $\Delta _{\text{eff}}$, the
Doppler energy shift [Eq. (\ref{DES})] is obtained.
Thus the SC gap of the occupied states ($E<0$) are spin-dependent, which is 
\begin{equation}
\Delta _{\sigma }=\Delta _{\sigma ,-}=\Delta _{\text{eff}}\left( 1-\sigma
D\right)
\end{equation}%
as shown in Fig. \ref{setup}(c).

\section{Scattering matrix and Andreev bound states} \label{App_B}
In this section, the calculation details of Andreev reflection (AR) coefficients and energy level of Andreev bound states (ABSs) are given in the framework of the scattering matrix \cite{Beenakker1991,Beenakker1992,Davydova2022}, corresponding to Fig. \ref{setup}(d) and (e) in the main text.
\subsection{The wavefunctions} \label{App_B1}
In a basis $\Psi (x)=[\psi _{\uparrow }(x),\psi _{\uparrow }^{\dagger
}(x),\psi _{\downarrow }(x),\psi _{\downarrow }^{\dagger }(x)]^{T}$, the
Hamiltonian describes a Josephson junction with a length $d$ supporting nonreciprocal current is%
\begin{equation}
H_{JJ}=\int \Psi ^{\dagger }(x)\hat{H}_{BdG}(-i\partial _{x})\Psi (x)dx\text{%
,}  
\end{equation}%
with 
\begin{equation}
\hat{H}_{BdG}(-i\partial _{x})=\left( 
\begin{array}{cc}
\hat{H}_{BdG}^{\uparrow }(-i\partial _{x}) & 0 \\ 
0 & \hat{H}_{BdG}^{\downarrow }(-i\partial _{x})%
\end{array}%
\right)  \label{hblock}
\end{equation}%
where the gate-voltage and SC order parameter are spatially dependent, $%
V\left( x\right) =V_{\text{SC}}\Theta \left( x^{2}-dx\right) +V_{\text{TG}}\Theta \left(
dx-x^{2}\right) $ and $\Delta \left( x\right) =\Delta e^{i\varphi
_{\text{L}}}\Theta \left( -x\right) +\Delta e^{i\varphi _{\text{R}}}\Theta \left(
x-d\right) $. 
Here, $V_{\text{SC}}$ ($V_{\text{TG}}$) is super (tunneling) gates applied in the SC lead (normal) region and $\varphi _{\text{L}}$ ($\varphi _{\text{R}}$) is the
macroscopic phases of the left (right) SC lead. We define $\varphi =\varphi_{\text{R}}-\varphi _{\text{L}}$ as the phase difference between two SC leads.

The block diagonalized BdG Hamiltonian Eq. (\ref{hblock}) allows us to solve the scattering problem in each block separately. 
The electron wave function in the N region is 
\begin{equation}
\psi _{e,\sigma }^{N}=\frac{1}{\sqrt{\sigma v_{\sigma }}}\left( 
\begin{array}{c}
1 \\ 
0%
\end{array}%
\right) e^{i\sigma k_{e,\sigma }^{N}x}\text{,}
\end{equation}%
while its hole counterpart is 
\begin{equation}
\psi _{h,\sigma }^{N}=\frac{1}{\sqrt{\sigma v_{\sigma }}}\left( 
\begin{array}{c}
0 \\ 
1%
\end{array}%
\right) e^{-i\bar{\sigma}k_{h,\bar{\sigma}}^{N}x}\text{,}
\end{equation}%
with 
\begin{equation}
k_{\gamma ,\sigma }^{N}=\frac{k_{E}-\sigma k_{m}+\gamma k_{TG}}{1+\gamma
\sigma t}\text{,}
\end{equation}%
where $k_{E}=E/\hbar v_{0}$, $k_{TG}=V_{TG}/\hbar v_{0}$ and $\gamma =+1$ ($%
-1$)\ for the electron (hole) block. 
In the SC region, the wave function is
\begin{equation}
\psi _{\gamma ^{\prime },\sigma }^{SC}=\left( 
\begin{array}{c}
e^{i\varphi _{\alpha }} \\ 
\sqrt{\frac{\sigma +t}{\sigma -t}}e^{-i\gamma ^{\prime }\theta _{\sigma }}%
\end{array}%
\right) e^{ik_{\gamma ^{\prime },\sigma }^{SC}x}\text{,}
\end{equation}%
where%
\begin{widetext}
\begin{equation}
k_{\gamma ^{\prime },\sigma }^{SC}=\gamma ^{\prime }\frac{1}{1-t^{2}}\left[ 
\sqrt{\left( tk_{SC}-\sigma k_{E}+k_{m}\right) ^{2}-k_{\Delta _{\text{eff}%
}}^{2}}+s\sigma \left( k_{SC}-t\sigma k_{E}+tk_{m}\right) \right] \text{,}
\end{equation}%
\end{widetext}
and 
\begin{equation}
\cos \theta _{\sigma }=\frac{\sigma k_{E}-k_{m}-tk_{SC}}{k_{\Delta _{\text{%
eff}}}}\text{,}
\end{equation}%
where $k_{\Delta _{\text{eff}}}=\Delta _{\text{eff}}/\hbar v_{0}$, $%
k_{SC}=V_{SC}/\hbar v_{0},\ \gamma ^{\prime }=+1$ ($-1$)\ for electron-like (hole-like) quasiparticles.
\subsection{The scattering matrix} \label{App_B2}
The propagating matrix $s_{N}$ in N region is 
\begin{equation}
s_{N}=\left( 
\begin{array}{cc}
s_{0} & 0 \\ 
0 & s_{0}^{\ast }%
\end{array}%
\right) =\left( 
\begin{array}{cccc}
r & t^{\prime } & 0 & 0 \\ 
t & r^{\prime } & 0 & 0 \\ 
0 & 0 & r^{\ast } & t^{\prime \ast } \\ 
0 & 0 & t^{\ast } & r^{\prime \ast }%
\end{array}%
\right) \text{,}  \label{sn}
\end{equation}%
which is defined by connecting the incident wave $\psi _{in}=[\psi
_{e,\downarrow }^{N}\left( 0\right) ,\psi _{e,\uparrow }^{N}\left( d\right)
,\psi _{h,\downarrow }^{N}\left( 0\right) ,\psi _{h,\uparrow }^{N}\left(
d\right) ]^{T}$ to the the reflected/transmitted wave $\psi _{out}=[\psi
_{e,\uparrow }^{N}\left( 0\right) ,\psi _{e,\downarrow }^{N}\left( d\right)
,\psi _{h,\uparrow }^{N}\left( 0\right) ,\psi _{h,\downarrow }^{N}\left(
d\right) ]^{T}$ in the N region, $\psi _{out}=s_{N}\psi _{in}$. Note that,
because of the spin-momentum locking (helicity), backscattering is
prohibited, so we have $t^{\prime }=t=1$ and $r^{\prime }=r=0$ in Eq. (\ref%
{sn}).

Similarly, the scattering matrix connecting the wavefunctions in the AR process at the N-SC interfaces is 
\begin{equation}
s_{A}=\left( 
\begin{array}{cccc}
0 & 0 & r_{eh}^{\downarrow \uparrow } & 0 \\ 
0 & 0 & 0 & r_{eh}^{\uparrow \downarrow } \\ 
r_{he}^{\downarrow \uparrow } & 0 & 0 & 0 \\ 
0 & r_{he}^{\uparrow \downarrow } & 0 & 0%
\end{array}%
\right) \text{,}  \label{sa}
\end{equation}%
which satisfies $\psi _{in}=s_{A}\psi _{out}$. Note that at $x=0$, we have 
\begin{eqnarray}
r_{eh}^{\uparrow \downarrow } &=&r_{A}^{\uparrow }e^{i\varphi _{L}}\text{,}
\\
r_{he}^{\uparrow \downarrow } &=&r_{A}^{\downarrow }e^{-i\varphi _{L}}\text{,%
}
\end{eqnarray}%
while at $x=d$, we have%
\begin{eqnarray}
r_{he}^{\downarrow \uparrow } &=&r_{A}^{\uparrow }e^{-i\varphi
_{R}}e^{iq_{\uparrow }d}\text{,} \\
r_{eh}^{\downarrow \uparrow } &=&r_{A}^{\downarrow }e^{i\varphi
_{R}}e^{iq_{\downarrow }d}\text{.}
\end{eqnarray}%
where $q_{\sigma }=(k_{e,\sigma }^{N}+k_{h,\sigma }^{N})=2(E-\sigma m-\sigma
tV_{TG})/[\hbar v_{0}\left( 1-t^{2}\right) ]$ and 
\begin{equation}
r_{A,\sigma }=\sigma e^{-i\theta _{\sigma }}\text{,}  \label{ra}
\end{equation}%
with $\theta _{\sigma }=\cos ^{-1}(E/\Delta _{\text{eff}}-\sigma D)$ for $%
|E/\Delta _{\text{eff}}-\sigma D|\leq 1$ and $\theta _{\sigma }=-i\cosh
^{-1}(E/\Delta _{\text{eff}}-\sigma D)$ for $|E/\Delta _{\text{eff}}-\sigma
D|>1$. Thus Eq. (\ref{sa}) can be simplified as 
\[
s_{A}=\left( 
\begin{array}{cc}
0 & s_{A,\sigma } \\ 
s_{A,\bar{\sigma}}^{\ast } & 0%
\end{array}%
\right) \text{,} 
\]%
with 
\begin{equation}
s_{A,\sigma }=\left( 
\begin{array}{cc}
r_{A,\uparrow }e^{i\varphi _{L}} & 0 \\ 
0 & r_{A,\downarrow }e^{i\left( \varphi _{R}+q_{\downarrow }d\right) }%
\end{array}%
\right) \text{.}  \label{sas}
\end{equation} 
\subsection{Andreev bound states} \label{App_B3}
The ABS can be found through an equation \cite{Beenakker1991,Beenakker1992} 
\begin{equation}
\det \left( I-s_{A}s_{N}\right) =\det \left(
I-s_{A,\bar{\sigma}}^{*}s_{0}s_{A,\sigma}s_{0}^{\ast }\right) =0\text{.}  \label{det}
\end{equation}%
By substituting Eq. (\ref{sn}-\ref{sas}) into Eq. (\ref{det}), the ABS spectrum can be obtained in a transcendental equation 
\begin{equation}
E=\sigma \Delta _{\text{eff}}\left\{ D-\cos \left( \frac{\varphi ^{\prime }}{%
2}\right) sign\left[ \sin \left( \frac{\varphi ^{\prime }}{2}\right) \right]
\right\} \text{,}  \label{eabs2}
\end{equation}%
with $\varphi ^{\prime }=\varphi +q_{\sigma }d$. 
In the short junction limit ($d\ll l_{SC}$, $l_{SC}=\hbar v_{0}/\Delta $ is the superconducting coherent length), the ABS spectrum can be simplified as Eq. (\ref{eabs1}).

\section{The current phase relations and diode efficiency} \label{App_C}

In this section, the general form of current phase relations (CPRs) using the free energy are given, corresponding to the CPR and the diode efficiency in finite and zero temperatures exhibited in Fig. \ref{setup}(f) and (g).
In particular, the analytical formulas of both the CPR and the diode
efficiency are obtained at zero temperature by distinguishing the
contributions from ABSs and continuum states.
\subsection{General form} \label{App_C1}
The Josephson current flowing in the sample can be determined by the free energy \cite{Davydova2022,Scharf2021,Beenakker1991,Beenakker1992} which is 
\begin{equation}
I(\varphi )=\frac{2e}{\hbar }\partial _{\varphi }F(\varphi )\text{,}
\label{if}
\end{equation}%
where, without parity constraints \cite{Beenakker2013,Tkachov2015,Scharf2021}, the free energy is 
\begin{eqnarray}
    F&=&-\frac{1}{\beta }\sum_{E>0}\ln \left[ 2\cosh \frac{\beta E}{2}\right] \nonumber \\
&&+\int d^{2}r\frac{\left\vert \Delta \right\vert ^{2}}{\left\vert
g\right\vert }+\text{Tr}(H_{0})\text{.}
\end{eqnarray}
By neglecting the phase-independent terms in the second line, we obtain 
\begin{eqnarray}
F&=&-\frac{1}{\beta }\sum_{E>0}\ln \left[ 2\cosh \frac{\beta E}{2}\right] \nonumber \\ 
&=&-%
\frac{1}{\beta }\int_{0}^{\infty }dE\nu \left( E\right) \ln \left[ 2\cosh 
\frac{\beta E}{2}\right]
\end{eqnarray}
where the density of states is 
\begin{equation}
\nu \left( E\right) =-\frac{1}{\pi }\text{Im}\partial _{E}\ln \det \left(
I-s_{A}s_{N}\right) +const.\text{,}  \label{dos}
\end{equation}%
with the notion $const.$ denoting the phase-independent term. Here $\beta
^{-1}=k_{B}T$ ($k_{B}\equiv 1$ is the Boltzmann constant) and the sum over
energy $E$ contains both the contributions of ABSs and continuum states. 

Using Eq. (\ref{dos}), Eq. (\ref{if}) can be rewritten as 
\begin{widetext}
\begin{eqnarray}
I &=&\frac{2e}{\hbar }\partial _{\varphi }F  \nonumber \\
&=&\frac{2e}{\hbar }\frac{-1}{\beta }\partial _{\varphi }\int_{0}^{\infty
}dE\nu \left( E\right) \ln \left[ 2\cosh \frac{\beta E}{2}\right]   \nonumber
\\
&=&\frac{2e}{\hbar }\frac{-1}{\beta }\int_{0}^{\infty }dE\ln \left[ 2\cosh 
\frac{\beta E}{2}\right] \partial _{\varphi }\nu \left( E\right)   \nonumber
\\
&=&\frac{2e}{\hbar }\frac{-1}{\beta }\left( -\frac{1}{\pi }\right)
\int_{0}^{\infty }dE\ln \left[ 2\cosh \frac{\beta E}{2}\right] \partial
_{\varphi }\left[ \text{Im}\partial _{E}\ln \det \left( I-s_{A}s_{N}\right) %
\right]  \\
&=&\frac{2e}{\pi \hbar \beta }\int_{0}^{\infty }dE\ln \left[ 2\cosh \frac{%
\beta E}{2}\right] \partial _{E}\left[ \text{Im}\partial _{\varphi }\ln \det
\left( I-s_{A}s_{N}\right) \right]   \nonumber \\
&=&\frac{2e}{\pi \hbar \beta }\int_{0}^{\infty }dE\ln \left[ 2\cosh \frac{%
\beta E}{2}\right] \partial _{E}\left[ \text{Im}\partial _{\varphi }\ln \det
\left( I-s_{A}s_{N}\right) \right]   \nonumber \\
&=&\frac{e}{\pi \hbar \beta }\int_{-\infty }^{\infty }dE\ln \left[ 2\cosh 
\frac{\beta E}{2}\right] \partial _{E}\left[ \text{Im}\partial _{\varphi
}\ln \det \left( I-s_{A}s_{N}\right) \right]   \nonumber \\
&=&\frac{e}{\pi \hbar \beta }\left[ \ln \left[ 2\cosh \frac{\beta E}{2}%
\right] \left[ \text{Im}\partial _{\varphi }\ln \det \left(
I-s_{A}s_{N}\right) \right] \right] _{-\infty }^{\infty }  \nonumber \\
&&-\frac{e}{\pi \hbar \beta }\int_{-\infty }^{\infty }dE\left[ \text{Im}%
\partial _{\varphi }\ln \det \left( I-s_{A}s_{N}\right) \right] \partial
_{E}\ln \left[ 2\cosh \frac{\beta E}{2}\right] \text{.}  \nonumber
\end{eqnarray}%
\end{widetext}
and note that $\left[ \cdots %
\right] _{-\infty }^{\infty }=0$ and $\partial _{E}\ln \left[ 2\cosh \frac{%
\beta E}{2}\right] =\frac{\beta }{2}\tanh \frac{\beta E}{2}$. The equation
above reduces as%
\begin{equation}
I=-\frac{e}{2\pi \hbar }\int_{-\infty }^{\infty }dE\tanh \frac{\beta E}{2}%
\text{Im}\partial _{\varphi }\ln \det \left( \hat{I}_{4\times4}-s_{A}s_{N}\right) \text{.}
\end{equation}
By substituting the scattering matrices Eqs. (\ref{sn})-(\ref{sas}), we obtain
\begin{equation}
I=\frac{2e}{\beta \hbar }\text{Re}\sum_{n=0}^{\infty }\cot \left[ \cos
^{-1}\left( i\frac{\omega _{n}}{\Delta _{\text{eff}}}+D\right) -\frac{%
\varphi }{2}\right] \text{,}  \label{Igen}
\end{equation}%
which describes general CPR in arbitrary temperature and $D$ and the short-junction limit is considered.

\subsection{Contributions from ABSs and continuum states} \label{App_C2}

When $\left\vert D\right\vert <1$, the zero-temperature current can be decoupled as 
\begin{equation}
I_{\left\vert D\right\vert <1}(\varphi )=I_{\text{ABS}}(\varphi )+I_{\text{cont}}(\varphi )\text{,}
\label{Ci1}
\end{equation}%
where%
\begin{equation}
I_{\text{ABS}}(\varphi )=-\frac{e}{\hbar }\sum_{\sigma }\sum_{E>0}^{E<\Delta
_{\sigma }}\partial _{\varphi }E_{\sigma }  \label{ibon}
\end{equation}%
and%
\begin{equation}
I_{\text{cont}}(\varphi )=-\frac{e}{\hbar }\int_{E=\Delta _{\sigma
}}^{+\infty }dE\times E\partial _{\varphi }\nu \left( E\right)  \label{icon}
\end{equation}%
are the contributions from ABS and continuum states to the current respectively. By substituting Eq. (\ref{eabs1}) into Eq. (\ref{ibon}), we have Eq. (\ref{iabs})
After some calculations, we obtain the current from continuum states Eq. (\ref{icont}), which is phase-independent and directly induced by the Doppler energy shift.

However, the situation with $\left\vert D\right\vert >1$ is complicated because the contribution from ABS and continuum states can not be separated.
But, as we go back to Eq. (\ref{Igen}) and follow the process in Ref. \cite{Davydova2022}, we obtain the second line of Eq. (\ref{i1}).
When $\left\vert D\right\vert \gg 1$, the current is reduced as 
\begin{equation}
\lim_{\left\vert D\right\vert \gg 1}I_{2}=\frac{e\Delta _{\text{eff}}}{%
2\hbar }\frac{1}{\pi D} \cos \varphi \text{.}  \label{ilimt}
\end{equation}%

\subsection{The Diode Efficiency} \label{App_C3}

In the CPR [Eq. (\ref{i1})], when $\left\vert D\right\vert <1$, two
critical currents are 
\begin{eqnarray}
I_{c+} &=&\max \left[ i_{1},i_{2}\right] \text{,} \\
I_{c-} &=&|\min \left[ i_{1},i_{2}\right] |\text{,}
\end{eqnarray}%
where 
\begin{eqnarray}
i_{1} &=&I_{\left\vert D\right\vert <1}\left( \varphi =2\cos ^{-1}D\right) \text{,} \\
i_{2} &=&I_{\left\vert D\right\vert <1}\left( \varphi =\pi \right) \text{.}
\end{eqnarray}%
Thus, by definition, the diode efficiency is 
\begin{equation}
\eta _{\left\vert D\right\vert <1}=sign\left( D\right) \frac{4\left\vert D\right\vert /\pi +\sqrt{%
1-D^{2}}-1}{\sqrt{1-D^{2}}+1}\text{,}  \label{Ceta1}
\end{equation}%
where $\sin \left( \arccos x\right) =\sqrt{1-x^{2}}$ is used. 

When $%
\left\vert D\right\vert >1$, two critical currents are%
\begin{eqnarray}
I_{c+} &=&\max \left[ i_{1},i_{2}\right] \text{,} \\
I_{c-} &=&|\min \left[ i_{1},i_{2}\right] |\text{,}
\end{eqnarray}%
where 
\begin{eqnarray}
i_{1} &=&I_{\left\vert D\right\vert \geq1}\left( \varphi =0\right) \text{,} \\
i_{2} &=&I_{\left\vert D\right\vert \geq1}\left( \varphi =\pi \right) \text{.}
\end{eqnarray}%
Then the JDE efficiency 
\begin{equation}
\eta _{\left\vert D\right\vert \geq1}=sign\left( D\right) \frac{2\left( \left\vert D\right\vert -\sqrt{%
D^{2}-1}\right) -\cot ^{-1}\sqrt{D^{2}-1}}{\cot ^{-1}\sqrt{D^{2}-1}}\text{.}
\label{eta2}
\end{equation}%

Since the cosinoidal CPR in Eq. (\ref{ilimt}), the diode effect vanishes at $%
\left\vert D\right\vert \gg 1$, i.e. 
\begin{equation}
\lim_{\left\vert D\right\vert \gg 1}\eta=0\text{.}
\end{equation}

\begin{figure}[t]
\centering \includegraphics[width=0.4\textwidth]{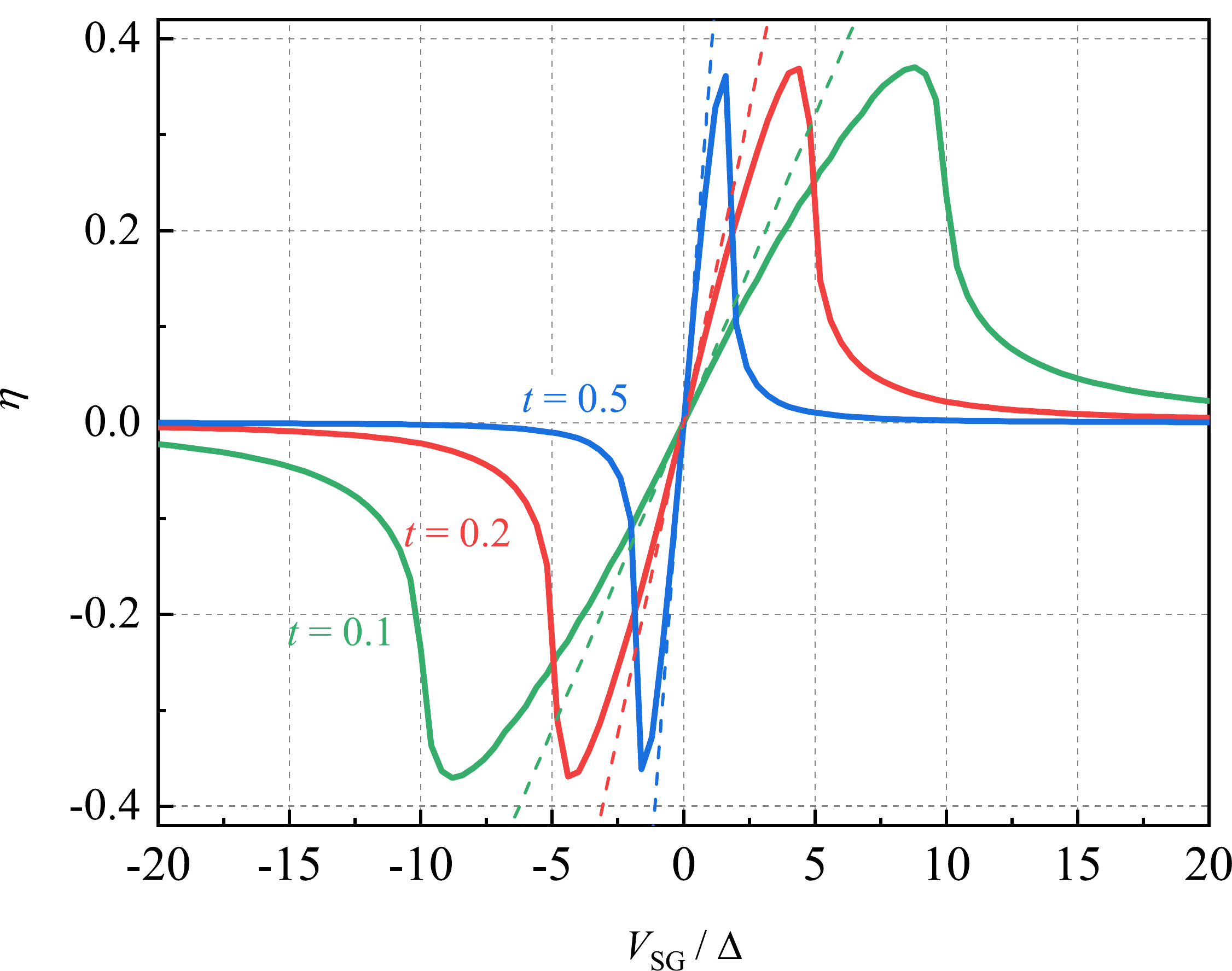}
\caption{The dependence of the diode efficiencies on super gate $V_{SC}$ with different strength of asymmetry. 
We choose $\hbar v_{0}=1=\Delta $ and $t=v_{t}/v_{0}=0.1$. For small $V_{SC}$, the diode efficiencies increase linearly via $V_{SC}$.}
\label{etaVSG}
\end{figure}

The dependence of diode efficiency with respect to the Doppler energy shift is exhibited in Fig. \ref{setup}(g).
Besides, the strength of asymmetry in a system can be estimated through the function diode efficiency with respect to the super gate $V_{SC}$ as shown in Fig. \ref{etaVSG}, where the $\eta$s linearly increase for small gate voltage with a slope relating to the strength of asymmetry.

\section{Topological Josephson diode} \label{App_D}

In this appendix and the following, we discuss two types of systems supporting the FEJD. 
One is the topological Josephson diode first proposed in this work and the others are based on semiconductors extensively operated in current experiments. 

Let's begin with the first case, including (i) a single-edge Josephson diode, (ii) a double-edge-interference Josephson diode, and (iii) the field-effect Josephson diode in a finite-length junction. 
The results are exhibited in Fig. \ref{topjd}(a) and (b).

\subsection{The model} \label{App_D1}

In a topological Josephson diode, the nonreciprocal current is carried by the anisotropic helical edge states, whose Hamiltonian of the spin-$\sigma $ mode at the $\text{s}$
edge is \cite{Fu2022}
\begin{equation}
H_{BdG}^{\text{s},\sigma }=\int \Psi _{\text{s},\sigma }^{\dagger }(x)\hat{H}%
_{BdG}^{\text{s},\sigma }(x)\Psi _{\text{s},\sigma }(x)dx\text{,}
\end{equation}%
where the basis is $\Psi _{\text{s},\sigma }(x)=[\psi _{\text{s},\sigma }(x),\psi _{\text{s},\bar{%
\sigma}}^{\dagger }(x)]^{T}$, and 
\begin{equation}
\hat{H}_{BdG}^{\text{s},\sigma }(x)=\left( 
\begin{array}{cc}
h_{e}^{\text{s},\sigma }(x) & \sigma \Delta \\ 
\sigma \Delta ^{\dagger } & -[h_{e}^{\text{s},\bar{\sigma}}(-x)]^{\ast }%
\end{array}%
\right) \text{,}
\end{equation}%
where $h_{e}^{\text{s},\sigma }(x)=s(\sigma \hbar v_{0}+\hbar v_{t})\left(
-i\partial _{x}\right) -V^{\text{s}}\left( x\right) +m_{\text{s}}$ is the Hamiltonian of
AHES and $V^{\text{s}}\left( x\right) =V_{SC}^{\text{s}}\Theta \left( x^{2}-dx\right)
+V_{TF}^{\text{s}}\Theta \left( dx-x^{2}\right) $ are edge-dependent.

It is worth noting that $m_{s}$ can be induced by an in-plane Zeeman field aligned with the edge-state spin quantization axis \cite{Dolcini2015} or by an (orbital) out-off plane magnetic flux inducing gauge invariant shift of momentum \cite{Tkachov2015}. 
However, the effect of the magnetic field is global and always opposite in two edges, $m_{\text{t}}=-m_{\text{b}}$, canceling the diode effect in the whole junction, even though the diode efficiency is measurable in an individual edge \cite{Dolcini2015}. 
Thus, we neglect the effect of the magnetic field by setting $m_{s}=0$ and the Doppler energy shift $ D\rightarrow sD^{\text{s}}=tV_{SC}^{s}/\Delta _{\text{eff}}$ is solely controlled by the $V_{SC}^{s}$ on each edge.
\subsection{A single-edge Junction}\label{App_D2}
The CPR, as well as the diode efficiency in a single edge, can be easily obtained by replacing the Doppler energy shift $D$ in the above sections as $sD^{\text{s}}=(tV_{SC}^{\text{s}}+sm_{\text{s}})/\Delta _{\text{eff}}$, which are edge-dependent. 
The results are listed as follows:

The CPR of a single edge is 
\begin{equation}
I^{\text{s}}\left( \varphi \right) =\left\{ 
\begin{array}{ll}
I^{\text{s}}_{\text{ABS}}( \varphi ) +I^{\text{s}}_{\text{cont}}( \varphi )  & \text{, } \vert
D^{\text{s}}\vert <1 \\ 
( e\Delta _{\text{eff}}/2\hbar) sign\left( D^{\text{s}}\right) \Gamma^{\text{s}} (
\varphi )  & \text{, } \vert D^{\text{s}}\vert \geq 1%
\end{array}%
\right. \text{,}  \label{i1edge}
\end{equation}%
and the corresponding diode efficiency is 
\begin{equation}
\eta^{\text{s}} =\frac{D^{\text{s}}}{|D^{\text{s}}|} \times \left\{ 
\begin{array}{ll}
\frac{4|D^{\text{s}}|/\pi +\sqrt{1-D^{\text{s}2}}-1}{\sqrt{1-D^{\text{s}2}}+1} & \text{, }|D^{\text{s}}|<1 \\ 
\frac{2(|D|-\sqrt{D^{\text{s}2}-1})-\cot ^{-1}\sqrt{D^{\text{s}2}-1}}{\cot ^{-1}\sqrt{D^{\text{s}2}-1%
}} & \text{, }|D^{\text{s}}|\geq1%
\end{array}%
\right. \text{.}  \label{eta1edge}
\end{equation}%

\subsection{Interference effect between two edge currents}\label{App_D3}

A QSHI-based Josephson junction is a natural platform for a superconducting
quantum interference device, where the total CPR at zero temperature is 
\begin{equation}
I^{\text{tot}}\left( \varphi \right) =\sum_{s=\text{t,b}}I^{\text{s}}\left(
\varphi \right) \text{,}  \label{Itot}
\end{equation}%
whose concrete form depends on the Doppler energy shift in each edge allowing local manipulation on two currents.

We first focus on the situation with $0\leq D^{\text{b}}<1$ and $D^{\text{t}%
}>0$. When $0\leq D^{\text{t}}<1$, Eq. (\ref{Itot}) is 
\begin{equation}
I^{\text{tot}}(\varphi )=\sum_{s=\text{t,b}}I_{1}^{\text{s}}(\varphi )\text{,%
}  \label{It1}
\end{equation}%
where the critical currents occur near the topologically protected zero-energy crossings in ABS spectra, which are 
\begin{equation}
I_{c+}=I^{\text{tot}}(\varphi _{c}^{\text{t}})\text{,}
\end{equation}%
and 
\begin{equation}
I_{c-}=|I^{\text{tot}}(2\pi -\varphi _{c}^{\text{b}})|\text{.}
\end{equation}%
The resulting diode efficiency is shown as the first line of Eq. \ref{etatot}.
Similarly, in the case of $D^{\text{t}}>1$, Eq. (\ref{Itot}) is 
\begin{equation}
I^{\text{tot}}(\varphi )=I_{2}^{\text{t}}(\varphi )+I_{1}^{\text{b}}(\varphi
)  \label{it2}
\end{equation}%
where the critical currents are 
\begin{eqnarray}
I_{c+} &=&I^{\text{tot}}(\pi )\text{,} \\
I_{c-} &=&|I^{\text{tot}}(2\pi -\varphi _{c}^{\text{b}})|\text{,}
\end{eqnarray}%
and the resulting diode efficiency is 
the second line of Eq. \ref{etatot}.

Additionally, when $D^{\text{t}}\gg 1$, since the top-edge current is negligible [Eq. (\ref{ilimt})], the total current is dominated by the bottom-edge one and thus the diode efficiency tends to be $\eta ^{\text{tot}}\rightarrow \eta _{1}^{\text{b}}$. 
Therefore when $0\leq D^{\text{b}}<1$ and $D^{\text{t}}>0$, the diode efficiency of the total Josephson current is shown as Eq. (\ref{etatot}), which takes an analogous form for arbitrary $D^{\text{t}}$ and $D^{\text{b}%
} $.

\begin{figure}[t]
\centering \includegraphics[width=0.45\textwidth]{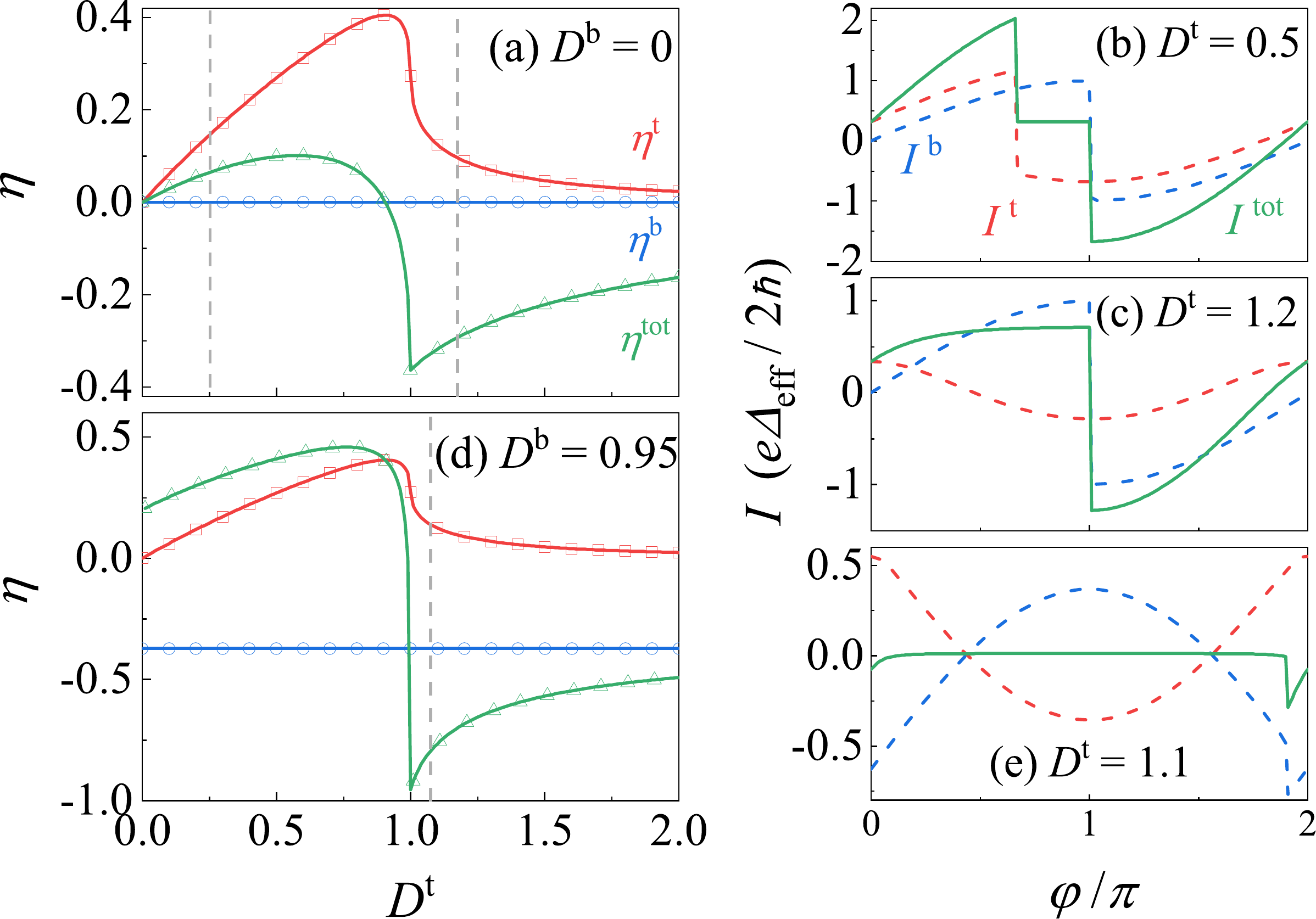}
\caption{The diode efficiencies of the top-edge (red), bottom-edge (blue),
and the total (green) current via the Doppler energy shift in the top edge ($%
D^{\text{t}}$) and the CPR. 
The Doppler energy shift in the bottom edge is (a) $D^{\text{%
b}}=0$ and (c) $D^{\text{b}}=0.95$ corresponding to the CPR in (b)/(c) with $D^{\text{t}}=0.5$/$D^{\text{t}}=1.2$ and (d) $D^{\text{t}}=1.2$,
respectively. Other parameters are chosen as Fig. \ref{etaVSG}. }
\label{inf}
\end{figure}

The Josephson diode efficiencies via $D^{\text{t}}$ and the CPRs are exhibited in Fig. \ref{inf}. 
The behaviors of $\eta ^{\text{t}}$ and $\eta ^{\text{b}}$ are fit with Eq. (\ref{eta1edge}), where the latter keeps zero due to $D^{\text{b}}=0$. 
As $D^{\text{t}}$ increases, the low diode efficiency of $\eta ^{\text{tot}}$ when $D^{\text{t}}<1$ suddenly inverts and becomes relatively measurable when $D^{\text{t}}>1$, while the efficiency in each edge remains low.

These phenomena are attributed to the interference between two currents and can be understood via the CPRs as shown in Fig. \ref{inf}. When $D^{\text{b}}=0$ and $D^{\text{t}}<1$, the CPRs of the two edge current are shown in Fig. \ref{inf}(b). 
Both the positive and opposite critical currents result from the constructive interference between two edge currents, which is approximate to be $I_{c+/c-}^{\text{tot}}\sim I_{c+/c-}^{\text{t}}+I_{c+/c-}^{\text{b}}$ in which $I_{c+}^{\text{b}}=I_{c-}^{\text{b}}$.
Thus, the resulting diode efficiency is suppressed since $\eta ^{\text{tot}}\sim (I_{c+}^{\text{t}}-I_{c-}^{\text{t}})/(I_{c+}^{\text{t}}+I_{c-}^{\text{t}}+2I_{c+}^{\text{b}})<\eta ^{\text{t}}$.

When $D^{\text{t}}>1$, the CPRs are shown in Fig. \ref{inf}(c). 
Because of the nearly cosinoidal behavior of the top-edge current [Eq. (\ref{i1edge})], the positive (negative) critical current of $I^{\text{tot}}\left( \varphi \right) $ occurs as $\varphi \sim \pi $, causing a destructive (constructive) interference between two currents, i.e. $I_{c+/c-}^{\text{tot}}\sim I_{c+}^{\text{b}}+(-)I_{2}^{t}\left( \pi \right) $ [$I_{2}^{t}\left(\pi \right) <0$]. 
Thus, the resulting diode efficiency is $\eta ^{\text{tot}}=I_{2}^{t}\left( \pi \right) /I_{c+}^{\text{b}}$, which is measurable even through the diode effects of the edge current nearly vanish.

Now, by combining the effect of the interferometer and Doppler energy shift in the bottom edge, the diode efficiency is highly optimized in the region of $D^{\text{t}}>1$, reaching up to $\eta ^{\text{tot}}\sim 90\%$ when $D^{\text{b,t}}\sim 1$ as shown in Fig. \ref{inf}(c).

Generally, a fully polarized diode with $|\eta ^{\text{tot}}|=1$ is impossible. 
This is restricted by the relation $\int_{0}^{2\pi }d\varphi I^{s}\left( \varphi \right) =0$, which indicates that when the critical current in one direction is zero, its opposite-direction counterpart must vanish. 
As shown in Fig. \ref{inf}(d), even though the positive critical current is nearly zero in a wide range of $\varphi $, the negative critical current only survives in a tiny range of $\varphi $.

Note that when $|D^{\text{t/b}}|\rightarrow +\infty $, the corresponding current vanished, $I^{\text{t/b}}\rightarrow 0$ and the diode reduces to $%
\eta _{\text{tot}}\rightarrow \eta _{1}^{\text{b/t}}$. 
Thus, in this situation, the current in a double junction is always larger than the one in a single junction. 

\begin{figure}[t]
\centering \includegraphics[width=0.45\textwidth]{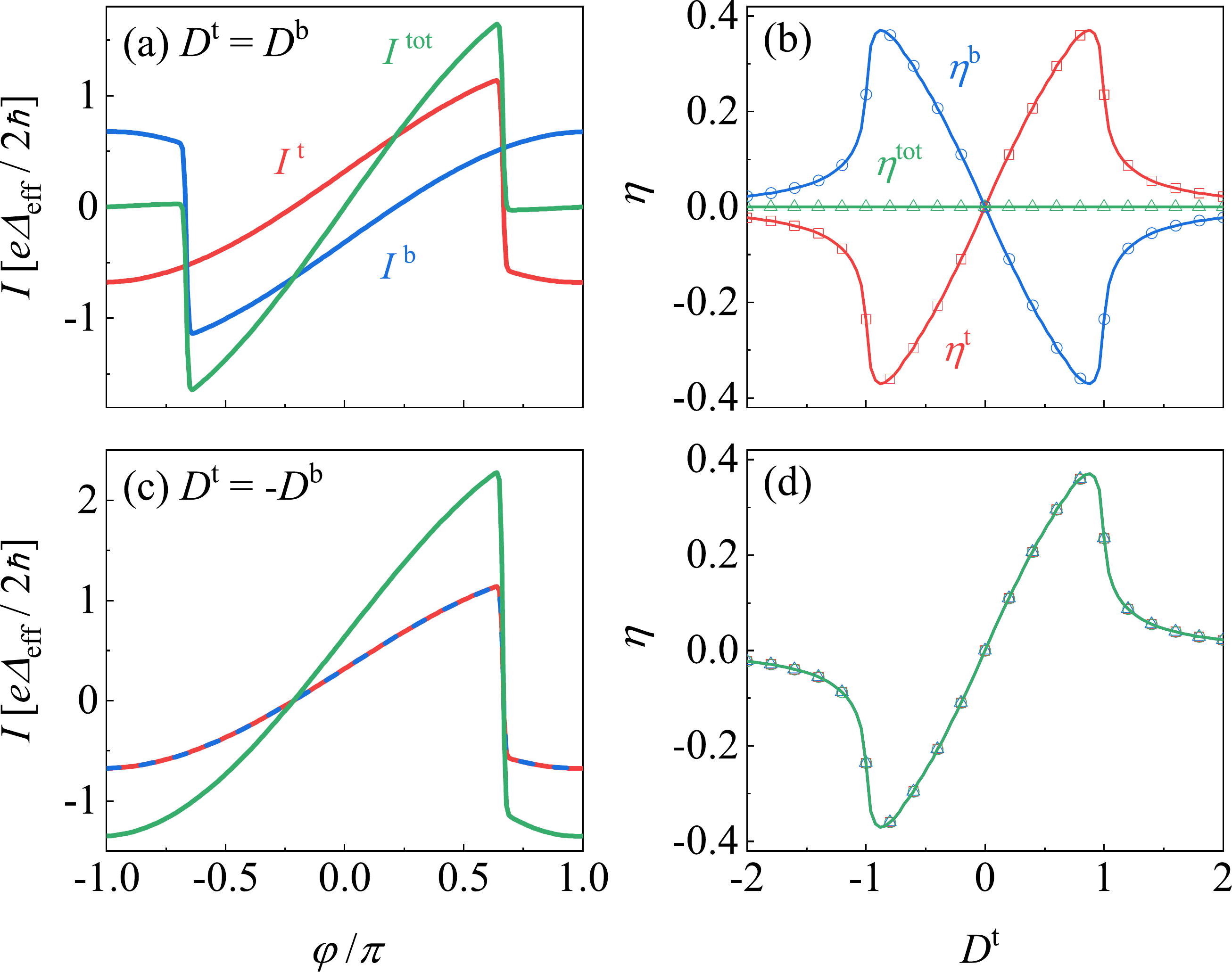}
\caption{
The top-edge (red), bottom-edge (blue),
and the total (green) CPR (left panel) and the diode efficiencies of (right panel) when two Doppler energy shifts are (a, b) identical and (c, d) opposite to each other.}
\label{inf2}
\end{figure}

It is worth noting that two special cases where the Doppler energy shift in each edge is identical and opposite to each other, as shown in Fig. (\ref{inf2}). 
In the former, no diode effect is expected, even though the supercurrent is asymmetric on each edge. 
This is the reason why no Josephson diode effect is achieved by applying a magnetic field \cite{Tkachov2015,Dolcini2015}.
On the other hand, in the latter case, the total diode effect behavior with the same tendency as the ones in each edge. This suggests that a topological FEJD can be realized in a time-reversal broken QSHI tunning a transverse electrical field \cite{Fu2022}

\subsection{Numerical simulations for field-effect topological Josephson diode} \label{App_D4}

So far, only the super-gate-controlled topological Josephson diode is discussed due to the negligible effect of the tunneling gate with the assumption of $d=0$. 
For a junction of finite length, a field-effect topological Josephson diode is expected in the numerical simulations, where the tunneling gate $V_{TG}$ can switch the diode on and off and even reverse the diode polarity.
\begin{figure}[t]
\centering \includegraphics[width=0.45\textwidth]{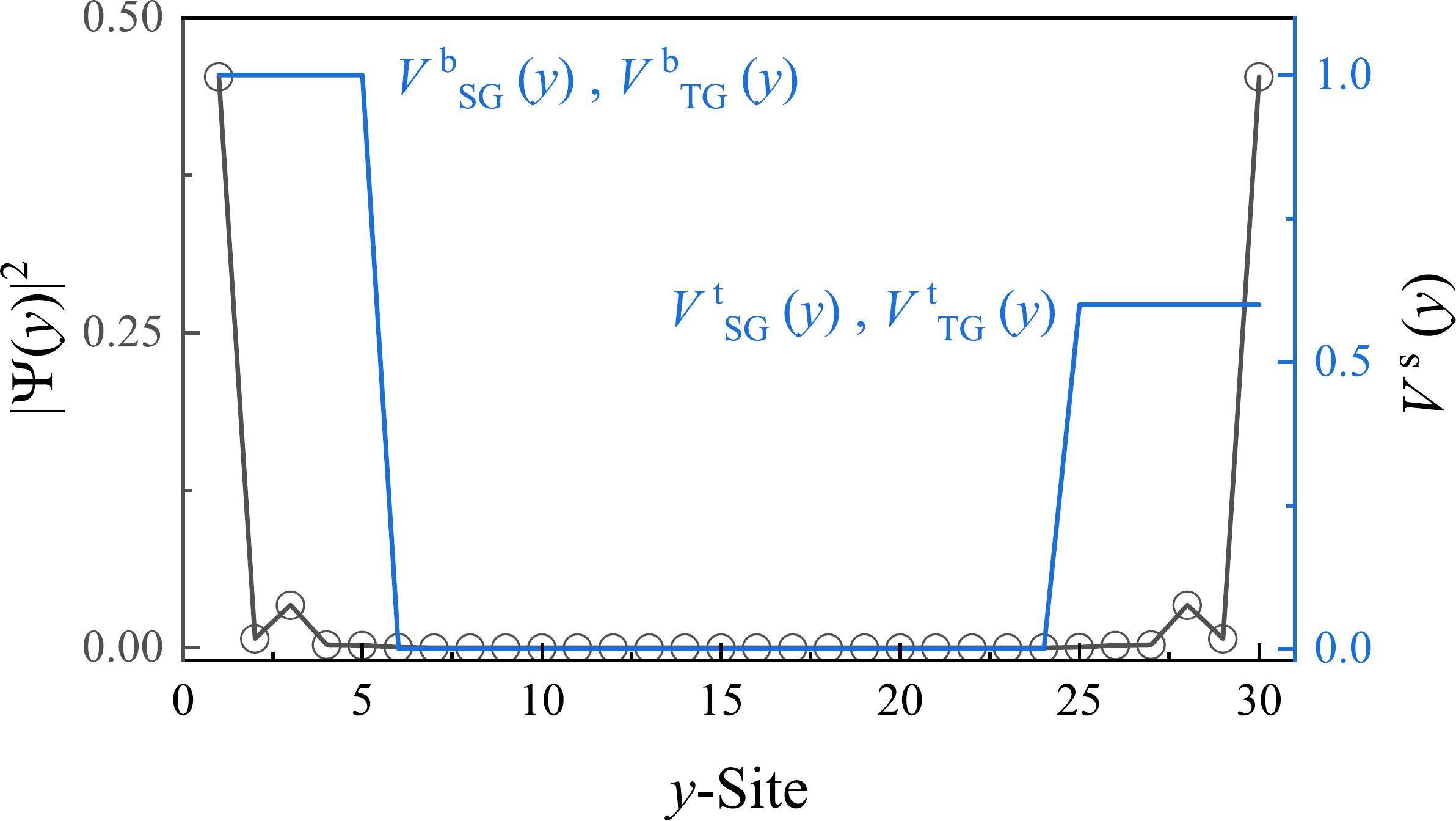}
\caption{A schematic for the distribution of $V(y)$ (in an arbitrary unit)
and the edge-state wave function along $y$-direction. }
\label{distri}
\end{figure}
\subsubsection{Tight-binding model and lattice Green's function}

By substituting $k_{x,y}$ and $k_{x,y}^{2}$ with $a^{-1}\sin (k_{x,y}a)$ and  $2a^{-2}[1-\cos (k_{x,y}a)]$ and using Fourier transformation, Eq. (\ref{hqshi}) can be rewritten in a square lattice. 
The Hamiltonian of the Josephson junction is 
\begin{equation}
H_{J}=\sum_{i=0}^{d}H_{N}^{latt}+\sum_{i<0}H_{L}^{latt}+%
\sum_{i>d}H_{R}^{latt}+H_{C}\text{,}  \label{HJ}
\end{equation}%
where 
\begin{equation}
H_{N}^{latt}=\sum_{j}^{L_{y}}C_{i,j}^{\dagger }\hat{h}%
_{0}C_{i,j}+C_{i+1,j}^{\dagger }\hat{h}_{x}C_{i,j}+C_{i,j+1}^{\dagger }\hat{h%
}_{y}C_{i,j}+h.c.\text{,}  \label{HNlatt}
\end{equation}%
and 
\begin{equation}
H_{L/R}^{latt}=\sum_{i<0,i>d}H_{N}^{latt}+\sum_{j,\sigma }\sigma \Delta
e^{i\varphi _{L/R}}C_{i,j,\sigma }C_{i,j,\sigma }+h.c.
\end{equation}%
are the Hamiltonian in the N region and left/right SC leads, respectively.
And 
\begin{equation}
H_{C}=\sum_{i=0/d,j}\hat{t}_{c}C_{i+1,j}^{\dagger }C_{i,j}+h.c.\text{,}
\label{HC}
\end{equation}%
couples the left/right SC to the central region. Here, $C_{i,j}=%
\{c_{i,j},c_{i,j}^{\dagger }\}^{T}$ with $c_{i,j}=\{c_{i,j,s,\uparrow
},c_{i,j,p,\uparrow },c_{i,j,s,\downarrow },c_{i,j,p,\downarrow }\}^{T}$ and 
$c_{i,j,s/p,\uparrow /\downarrow }$ ($c_{i,j,s/p,\uparrow /\downarrow
}^{\dagger }$) is the annihilation (creation) operator of $s$/$p$\ orbit
electron at site $\{i,j\}$ with spin $\uparrow $/$\downarrow $ and the
notion $\sigma =\pm 1$ represents spin up and down.

The components are 
\begin{eqnarray}
h_{0} &=&\xi _{z}\otimes h_{0}^{e} \\
h_{x} &=&\xi _{z}\otimes h_{x,1}^{e}+\xi _{0}\otimes h_{x,2}^{e} \\
h_{y} &=&\xi _{z}\otimes h_{y}^{e} \\
\hat{t}_{c} &=&-t_{c}\xi _{z}\otimes \sigma _{0}\otimes \tau _{0}
\end{eqnarray}%
where $\xi _{0}$ ($\xi _{z}$) is the unit (the third Pauli) matrix in Nambu
space, and 
\begin{eqnarray}
h_{0}^{e} &=&\left( C_0-\frac{4}{a^{2}}C_1\right) \sigma _{0}\tau _{0} +\left(
M_{\gamma }-\frac{4}{a^{2}}B_{\gamma }\right) \sigma _{0}\tau _{z} \nonumber \\
&&+M_{\alpha
}\sigma _{z}\tau _{0}+M_{\beta }\sigma _{z}\tau _{z}+V\left( x,y\right) \\
h_{x,1}^{e} &=&-\frac{1}{a^{2}}\left( C_1\sigma _{0}\tau _{0}-B_{0}\sigma
_{0}\tau _{z}\right) \\
h_{x,2}^{e} &=&-i\frac{1}{2a}\left( A\sigma _{x}\tau _{x}-A^{\prime }\sigma
_{y}\tau _{y}\right) \\
h_{y}^{e} &=&-\frac{1}{a^{2}}\left( C_1\sigma _{0}\tau _{0}-B_{0}\sigma
_{0}\tau _{z}\right) \nonumber \\
&&-i\frac{1}{2a}\left( A\sigma _{y}\tau
_{x}+A^{\prime }\sigma _{x}\tau _{y}\right) \\
t_{c} &=&\frac{4}{a^{2}}B_{\gamma }
\end{eqnarray}

In the numerical calculations we choose $M_{\beta }=0$ for simplification and the Fermi level is controlled by $V\left( x,y\right) $, which is spatially dependent, 
\begin{eqnarray}
V\left( x,y\right) &=&[V_{SC}^{\text{t}}\Theta \left( x^{2}-dx\right)
+V_{TG}^{\text{t}}\Theta \left( dx-x^{2}\right) ]\nonumber \\
&& \times \Theta \left( \lambda
_{y}y-y^{2}\right)  \nonumber \\
&&+[V_{SC}^{\text{b}}\Theta \left( x^{2}-dx\right) +V_{TG}^{\text{b}} \\
&& \times \Theta
\left( dx-x^{2}\right) ]\Theta \left[ \left( L_{y}-y\right) \left( y+\lambda
_{y}-L_{y}\right) \right] \nonumber \text{.}
\end{eqnarray}
The distribution of $V\left( x,y\right) $ along $y$-direction is shown in Fig. \ref{distri} compared with the wave function of the edge states. 

With the discussion above, using lattice Green's function method, the Josephson current through column $l$ in the N region is \cite{Fu2019,Xu2018} 
\begin{equation}
J=\frac{1}{h}\int_{-\infty }^{\infty }Tr[\hat{h}_{x}^{\dagger }\hat{e}%
G_{l,l-1}^{<}-\hat{e}\hat{h}_{x}G_{l-1,l}^{<}]dE,
\end{equation}%
where $\hat{e}=-e\xi _{z}\sigma _{0}\tau _{0}$ is the charge matrix. 
In equilibrium, the lesser-than-Green's function is calculated by $G^{<}=f(E)[G^{a}-G^{r}]$ where $f(E)$ is the Fermi-Dirac distribution function. 
The retarded Green's function is
\begin{equation}
G^{r}(E)=\frac{1}{E-H_{x}-\Sigma _{L}^{r}(E)-\Sigma _{R}^{r}(E)},
\end{equation}%
and $G^{a}=[G^{r}(E)]^{\dagger }$, where $H_{x}=\sum_{j}^{L_{y}}C_{i,j}^{%
\dagger }\hat{h}_{0}C_{i,j}+C_{i,j+1}^{\dagger }\hat{h}_{y}C_{i,j}+h.c.$ is the Hamiltonian of an individual column and the retarded self-energy $\Sigma_{L/R}^{r}\left( E\right) $ representing the coupling with left/right SC lead can be calculated numerically by the recursive method.

To understand the behavior of Josephson current, one can calculate ABS spectra through Green's function technique numerically. 
The energies of ABS levels can be located by searching the peaks of particle density within the SC gap at column $l$ 
\begin{equation}
\rho _{l}=-\frac{1}{\pi }\text{Im}( \text{Tr}G_{l,l}^{r})  \label{ABS}
\end{equation}%
at a given phase difference $\varphi$.

\subsubsection{Remarks on the numerical results}

As is observed in Eq. (\ref{hqshi}), the effects of TRS-breaking term on the QSHI include:

(i) the asymmetry to the helical edge states through the terms proportional
to $A^{\prime }$,

(ii) the $M_{\beta }$ term, which is not related to the band topology but
induces a spin-dependent shifting to the energy spectrum like a Zeeman
field, and

(iii) the $M_{\alpha }$ term, which modifies the bulk band gaps and can finally, drive the system to be in a QAHI phase when $\left\vert M_{\alpha
}\right\vert >M_{\gamma }$.


\begin{figure}[t]
\centering \includegraphics[width=0.45\textwidth]{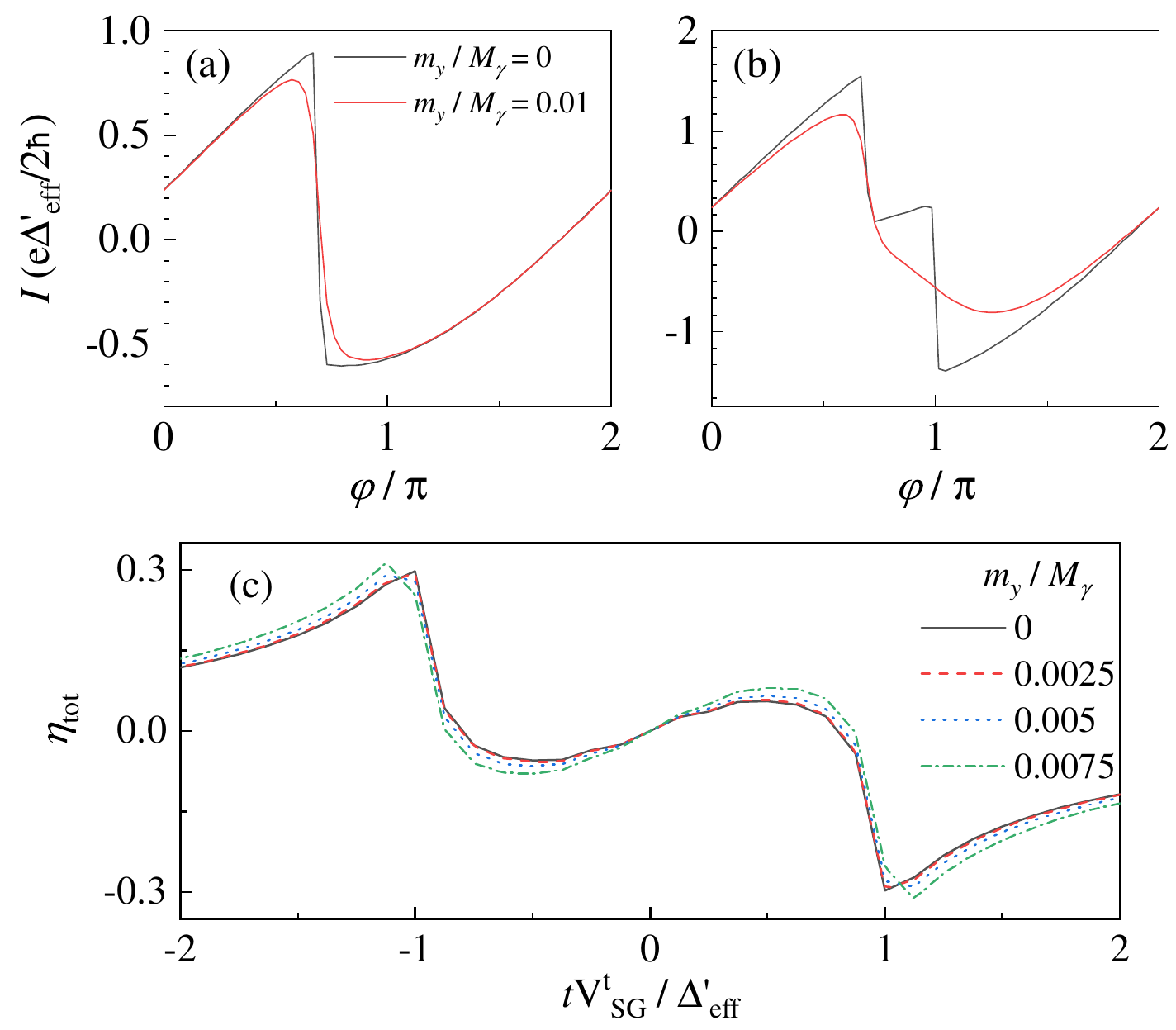}
\caption{The effect of $m_{y}$ on (a) single-edge CPR (top edge), (b) the total CPR and (c) the gate-controlled diode efficiency.
$m_{y}$ reduces the single-edge diode efficiency (a) due to backscattering but enhances the total diode efficiency due to the interference. 
In (a), $\eta_{\text{t}} = 0.19$ and $0.14$ for $m_y/M_\gamma=0$ and $0.01$, respectively.  
In (b), $\eta_{\text{tol}} = 0.05$ and $0.17$ for $m_y/M_\gamma=0$ and $0.01$, respectively.  
}
\label{my}
\end{figure}

Based on the observation of the numerical results, the three effects on the Josephson diode are concluded below.

Effect (i) leads to the gate-controlled Josephson diode which is the main claim of this work.

Effect (ii) leads to a gate-free diode effect on each edge [Fig. \ref{mb}(a)], while in the whole system, the diode effect vanishes because of the compensation between two edges [Fig. \ref{mb}(b)]. 
The reason is the Doppler energy shift depends on $M_{\beta }$ via $m_{\text{s}}=$s$M_{\beta }$. As a result, when $V_{SC}^{\text{s}}=0$, we obtain $D^{\text{t}}=-D^{\text{b}}$, and thus $\eta ^{\text{t}}=-\eta ^{\text{b}}$ and $\eta ^{\text{tot}}=0$. 
One of the motivations of
this work is the complementary effect between the electrical potential $tV_{SC}^{\text{s}}$ and the Zeeman field $m_{\text{s}}$.

Effect (iii) is dual. 
First, the Josephson diode vanishes when the system is driven to be a QAHI, which is beyond this work due to the destruction of helical edge states. 
Besides, we observe that edge-state superconducting gap $\Delta _{\text{edge}}$ depends on both the bulk-state superconducting gap $ \Delta _{\text{bulk}}$ as expected and bulk-state band gap controlled by $M_{\alpha }$ which is reported in Ref. \cite{Tkachov2019a}. 
Since an increasing $M_{\alpha }$ reduces $\Delta _{\text{edge}}$, a lower gate voltage is required to obtain the maximum value of the diode efficiency on each edge as shown in Fig. \ref{mb}.

On the other hand, it is possible to induce a gap in the edge spectrum by $H_{p}=m_y\sigma_y\tau_0$ when the TRS is broken. This couples the left- and right-mover in the edge leading to a back-scattering that smoothens the skewed CPR and thus reduces the diode efficiency in one edge as shown in Fig. \ref{my}(a). 
However, because of the interference, the total diode efficiency is enhanced [Fig. \ref{my}(b) and (c)].

\begin{figure}[t]
 \centering 
 \includegraphics[width=0.45\textwidth]{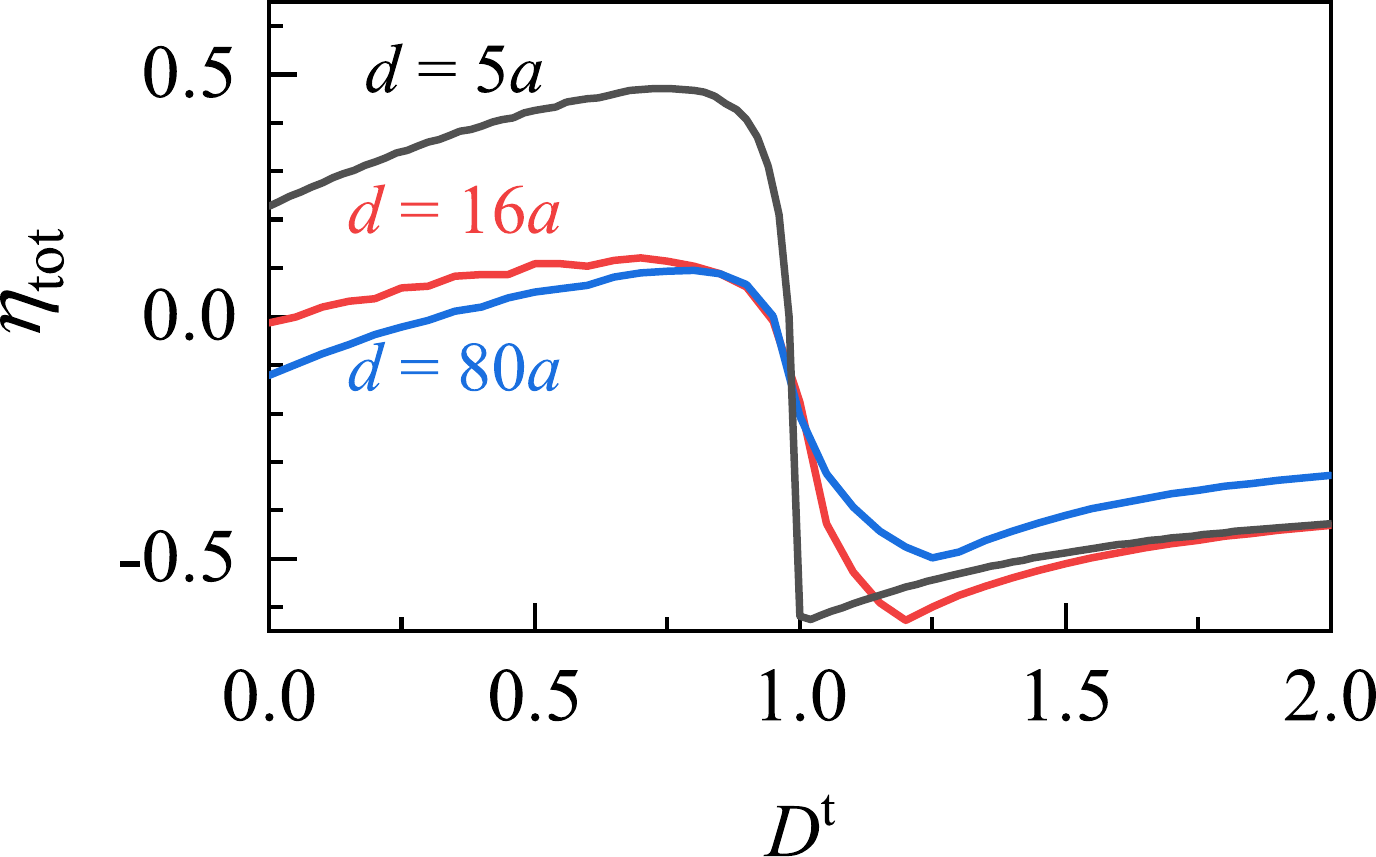}
 \caption{
 (a) Numerical calculations of $\eta_{\text{tot}}$ via $D^{\text{t}}$ in a tight-binding model. 
 The black line: result obtained in a short junction limit with $d=5a\ll l_{SC}\sim 50a$ and $D^{\text{b}}=-D^{\text{t}}$, slightly deviates from the Eq. (\ref{etatot}) the results obtained from the analysis results in Fig. \ref{inf2}(c). 
 The red and blue line: the results obtained in the intermediate-length $d=16a$ and long-junction limit $d=80a$. 
 In these two cases, to obtain a high diode efficiency the TG is fine-tuned as  $V_{TG}^{\text{t}}=1.9V_{TG}^{\text{t}}$ and  $V_{TG}^{\text{t}}=0.8V_{TG}^{\text{t}}$, respectively.}
 \label{num}
\end{figure}

When the junction length increases, as shown in Fig. \ref{num}, even though the short-junction condition preserves ($d=d_{0}/5\ll l_{SC}$), the behavior of $\eta ^{\text{tot}}$ via $D^{s}$ deviates from the Eq. (\ref{etatot}), but a relatively high-efficiency diode keeps when the gate is finely tuned. 
Such a deviation may be attributed to the additional phase accumulated $\varphi _{K}^{s}$ by the non-zero center-of-mass velocity, which reduces the interference effect and thus suppresses the diode efficiency. 

However, the phase difference between two interfering currents can be controlled by the TG, resulting in a field-effect Josephson diode transistor.
This manifests in oscillating diode efficiency with respect to the gate at the top edge as shown in Fig. \ref{topjd}(b), with an oscillating frequency increasing as the junction length grows.
In the diode transistor, the diode effect can be switched on with an efficiency up to $60\%$ when $D^{\text{b}}>1$, and the polarity can be rapidly reversed when $D^{\text{b}}>1$.
\begin{figure*}[t]
\centering \includegraphics[width=0.8\textwidth]{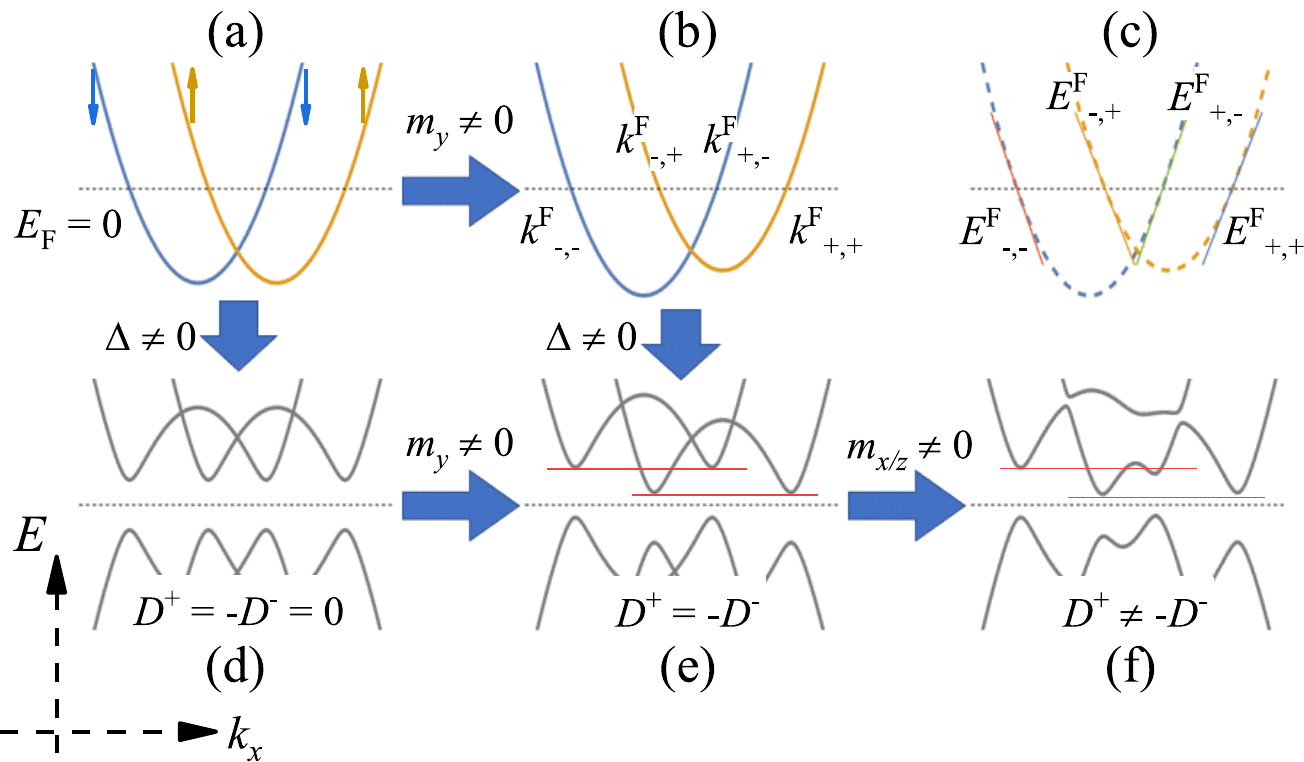}
\caption{ The schematic evolution of the dispersion of Eq. (\ref{h2deg}) and Eq. (\ref{EGBdG}) with $t_{0}a^2=1$, $\alpha_0=1$, $\varphi_{so}=\pi/2$, $V=0.5$, $k_{y}=0$ and different combination of $\{ m_{y},m_{x},m_{z},\Delta \}$.
The Fermi level ($E_{F}=0$) is represented as the black dashed line in each figure.
The spin orientation is defined with respect to the spin-orbit field with a rotation [see Eq. (\ref{h0rot})]. 
(a) No magnetic field or SC proximity effect. 
(b) A magnetic field along the direction of the spin-orbit field is applied, $m_y=0.1$. 
The Fermi wavenumber $k_{\alpha,\sigma}$ is obtained using Eq. (\ref{kF}). 
(c) A comparative plot between (b) (dashed line) and the effective Hamiltonian Eq. (\ref{hF}) around the Fermi wavenumber $k_{\alpha,\sigma}$ (solid line).
(d) The dispersion with superconducting pairing ($\Delta=0.2$). There is no Doppler energy shift ($D^{+}=-D^{-}=0$).
(e) The dispersion with superconducting pairing and $m_y=0.1$. The Doppler energy shift in two subbands is opposite ($D^{+}=-D^{-}\neq 0$). No diode effect is expected. 
(f) The superconducting dispersion with pairing $m_y=0.1$ and $m_{x/z}=0.2$, the magnetic field perpendicular to the spin-orbit field. 
It is possible to realize the Josephson diode due to the different magnitude of the Doppler energy shift in two subbands ($D^{+} \neq -D^{-}$).
}
\label{bandsemi}
\end{figure*}
\section{Semiconductor-based Josephson diode}\label{App_E}
In this appendix, we demonstrate that the framework of the ASMLSs can
explain the gate-tunability of the semiconductors Josephson diode
investigated in recent experiments \cite{Mazur2022}. 

We first exhibit that ASMLSs occur, when the Zeeman field is aligned with the spin-orbit field that couples to the momentum parallel to the current as illustrated in Fig. \ref{bandsemi}.
Then, the gate-tunability of the Josephson diode is numerically studied in the systems including a theoretical model of 1D limit and the experimentally concerned 2D nanoflakes.

The mechanism of the gate-tunability is explained by investigating the ABS and CPR exhibited in Fig. \ref{ABSCPREG}.
The diode efficiency is controlled by the gate through the electron density. 
In low (here is negative) gate voltage (Regime-I in Fig. \ref{ABSCPREG}), the system is an insulator and low electron density around the Fermi level fails to support the formation of Cooper pairs. 
Thus there is no Josephson current as well as the diode effect. 
For a medium gate voltage (Regime-II in Fig. \ref{ABSCPREG}), the system is a half metal, where one of the two channels is activated with high transmission probability, while the other channel keeps in the tunneling regime. 
The interference between two currents causes a tunneling-gate-control oscillating diode efficiency [Fig. (\ref{phimVSGVTG})] resembling double-junction interference in topological FEJD. 
Finally, two ballistic Josephson channels are activated by the high gate voltage (Regime-III in Fig. \ref{ABSCPREG}), and the diode oscillation is suppressed due to the constructive interference between the currents.

\subsection{Asymmetric spin-momentum-locking states around Fermi surface} \label{App_E2}

We consider a 2D electron gas (EG) with a strong spin-orbit interaction (SOI) and subjected to a Zeeman splitting. 
The Hamiltonian is shown as Eq. (\ref{h2deg}). 
Similarly, when taking the SC proximity effect into account, the BdG Hamiltonian of Eq. (\ref{h2deg}) is
\begin{equation}
h_{\text{2D-BdG}}(\bm{k})=\left(
\begin{array}{cc}
h_{\text{2D}}\left( \bm{k}\right)  & -i\Delta \sigma _{y} \\
i\Delta \sigma _{y} & -h_{\text{2D}}^{\ast }\left( -\bm{k}\right) 
\end{array}%
\right)\text{.}
\label{EGBdG}
\end{equation}

For the 1D limit of with $k_{y}=0$, pure RSOI ($\varphi _{so}=\pi /2$), and $m_{x}=m_{z}=0$, the Hamiltonian Eq. (\ref{h2deg}) is reduced as 
\begin{equation}
h_{\text{1D}}(k_x)=h_{\text{2D}}(k_x,k_y=0)\text{,}  \label{h1d}
\end{equation}%
whose dispersion is%
\begin{equation}
E_{\sigma }=(t_0a^2k_{x}^{2}-V)+\sigma \left( -\alpha _{R}ak_{x}+m_{y}\right) .
\label{Es}
\end{equation}%
As shown in Fig. \ref{bandsemi}, with the magnetic field aligned with the spin-orbit field direction, the band-touching nodes shift accompanied by a band tilting resulting in an ASML.

To illustrate within the framework of ASML, we rotate the spin polarization
axis along the spin-orbit field and the Eq. (\ref{h1d}) can be rewritten as 
\begin{eqnarray}
 h_{0}^{\prime }(k_{x})&=&U^{-1}h_{0}(k_{x})U\nonumber\\
 &=&(t_0a^2k_{x}^{2}-V)\sigma _{0}^{\prime
}-a\alpha _{R}k_{x}\sigma _{z}^{\prime }+m_{y}\sigma _{z}^{\prime } \text{,}
\label{h0rot}   
\end{eqnarray}
where 
\begin{equation}
U=\frac{1}{\sqrt{2}}\left( 
\begin{array}{cc}
1 & 1 \\ 
i & -i%
\end{array}%
\right) \text{.}
\end{equation}%
Assuming that the Fermi energy is zero, the Fermi wavenumbers are 
\begin{equation}
k_{\alpha =\pm 1,\sigma }^{F}=\alpha \kappa _{\sigma }+\sigma \frac{1}{2t_0a^2}%
\alpha _{R}\text{,}\label{kF}
\end{equation}%
with $2t_0a^2\kappa _{\sigma }=$\ $\sqrt{4t_0a^2\left( V-\sigma m_{y}\right) +(a\alpha
_{R})^{2}}$ and $k_{-,-}^{F}<k_{-,+}^{F}<k_{+,-}^{F}<k_{+,+}^{F}$ as shown in Fig. \ref{bandsemi}(b).
The Hamiltonian around Fermi surface can be written in the basis $\psi =(\psi_{+,\uparrow },\psi _{-,\downarrow },\psi _{-,\uparrow },\psi _{+,\downarrow})^{T}$ as%
\begin{equation}
h_{F}=\left( 
\begin{array}{cc}
h_{+} & 0 \\ 
0 & h_{-}%
\end{array}%
\right)\label{hF}
\end{equation}%
with 
\begin{equation}
h_{\alpha }=\left( 
\begin{array}{cc}
E_{\alpha ,+}^{F} & 0 \\ 
0 & E_{\bar{\alpha},-}^{F}%
\end{array}%
\right)  \label{ha}
\end{equation}%
where $\alpha =+$ ($-$) describe the states with a larger (smaller) magnitude of Fermi wavenumber and
\begin{equation}
E_{\alpha ,\sigma }^{F}=2t_0a^2\alpha \kappa _{\sigma }(q_{x}-k_{\alpha ,\sigma }^{F})
\end{equation}%
is obtained by expanding Eq. (\ref{Es}) around $k_{\alpha ,\sigma }^{F}$. 
A comparative plot in Fig. \ref{bandsemi}(c) verifies this effectiveness.

After a further rearrangement, Eq. (\ref{ha}) has the same form as Eq. (\ref{edge}) supporting ASMLSs, which is
\begin{equation}
h_{\alpha }\left( q_{x}\right) =\alpha \left( \hbar v_{t}^{\prime
}q_{x}\sigma _{0}+\hbar v_{0}^{\prime }q_{x}\sigma _{z}^{\prime }-\mu
^{\prime }\sigma _{0}+m^{\prime }\sigma _{z}^{\prime }\right) \text{,}
\end{equation}%
where $q_{x}=k_{x}-k_{\alpha ,\sigma }^{F}$ and 
\begin{eqnarray}
\hbar v_{t}^{\prime } &=&t_0a^2\left( \kappa _{+1}-\kappa _{-1}\right) \text{,} \\
\hbar v_{0}^{\prime } &=&t_0a^2\left( \kappa _{+1}+\kappa _{-1}\right) \text{,} \\
\mu ^{\prime } &=&t_0a^2(\kappa _{+}k_{\alpha ,+}^{F}-\kappa _{-}k_{-\alpha
,-}^{F})\text{,} \\
m^{\prime } &=&-t_0a^2\left( \kappa _{-}k_{-\alpha ,-}^{F}+\kappa _{+}k_{\alpha
,+}^{F}\right) \text{.}
\end{eqnarray}

However, when considering the SC proximity effect in Eq. (\ref{EGBdG}), the effect of the Doppler energy shift in two subbands always cancels each other, because 
\begin{equation}
D^{\alpha =+1}=\frac{t^{\prime }\mu ^{\prime }+m^{\prime }}{\Delta^{\prime }_{\text{eff}}}=-D^{-}\text{,}
\end{equation}%
as shown in Fig. \ref{bandsemi}(e) and no diode effect is expected.

This degeneracy is destroyed by the magnetic field component perpendicular
to the spin-orbit field shown in Fig. \ref{bandsemi}(f). 
Thus, both the Zeeman-field components parallel to and perpendicular to the spin-orbit field are necessary for the Josephson diode, which is verified in Fig. \ref{phimVSGVTG} through the numerical calculations below.

\subsection{Results and discussions}\label{App_E3}

\subsubsection{The tight-binding model}

The tight-binding version of Eq. (\ref{EGBdG}) is 
\begin{equation}
h_{EG}^{tb}=\sum_{\bm{r}_{i}}\left[ C_{\bm{r}_{i}}^{\dagger }\hat{h}%
_{0}C_{\bm{r}_{i}}+\sum_{\delta r}C_{\bm{r}_{i}+\delta \bm{r}}^{\dagger }\hat{h}%
_{\bm{r}}C_{\bm{r}_{i}}+h.c.\right] \text{,}  \label{tbeg}
\end{equation}%
where $\bm{r}_{i}=\left( x_{i},y_{i}\right) $ is the site index, $\delta \bm{r}=\left(\delta x_{i},\delta y_{i}\right) $ in the unit vector and $\hat{h}_{\bm{r}}=\left( \hat{h}_{x},\hat{h}_{y},\hat{h}_{z}\right) $ is the hopping matrix along three directions. 
The components in Eq. (\ref{tbeg}) are
\begin{eqnarray}
\hat{h}_{0} &=&\left( 6t_{0}-V\right) \xi _{z}\sigma _{0}+m_{x}\xi _{z}\sigma
_{x} \nonumber \\ 
&&+m_{y}\xi _{0}\sigma _{y}+m_{z}\xi _{z}\sigma _{z}-\Delta \xi _{y}\sigma
_{y}\text{,} \\
\hat{h}_{x} &=&-t_{0}\xi _{z}\sigma _{0}+\frac{i}{2a}\left( \alpha _{R}\xi
_{z}\sigma _{y}-\alpha _{D}\xi _{0}\sigma _{x}\right) \text{,} \\
\hat{h}_{y} &=&-t_{0}\xi _{z}\sigma _{0}-\frac{i}{2a}\left( \alpha _{R}\xi
_{0}\sigma _{x}-\alpha _{D}\xi _{z}\sigma _{y}\right) \text{,}
\end{eqnarray}%
combined with the lattice Green's function, the Josephson current and the ABS can be obtained.
The junction width is $W=30a$,
the Land\'e $g$-factors are isotropic $g_{x/z}=g_y$ and both RSOI and DSOI are considered.

\begin{figure}[t]
\centering \includegraphics[width=0.45\textwidth]{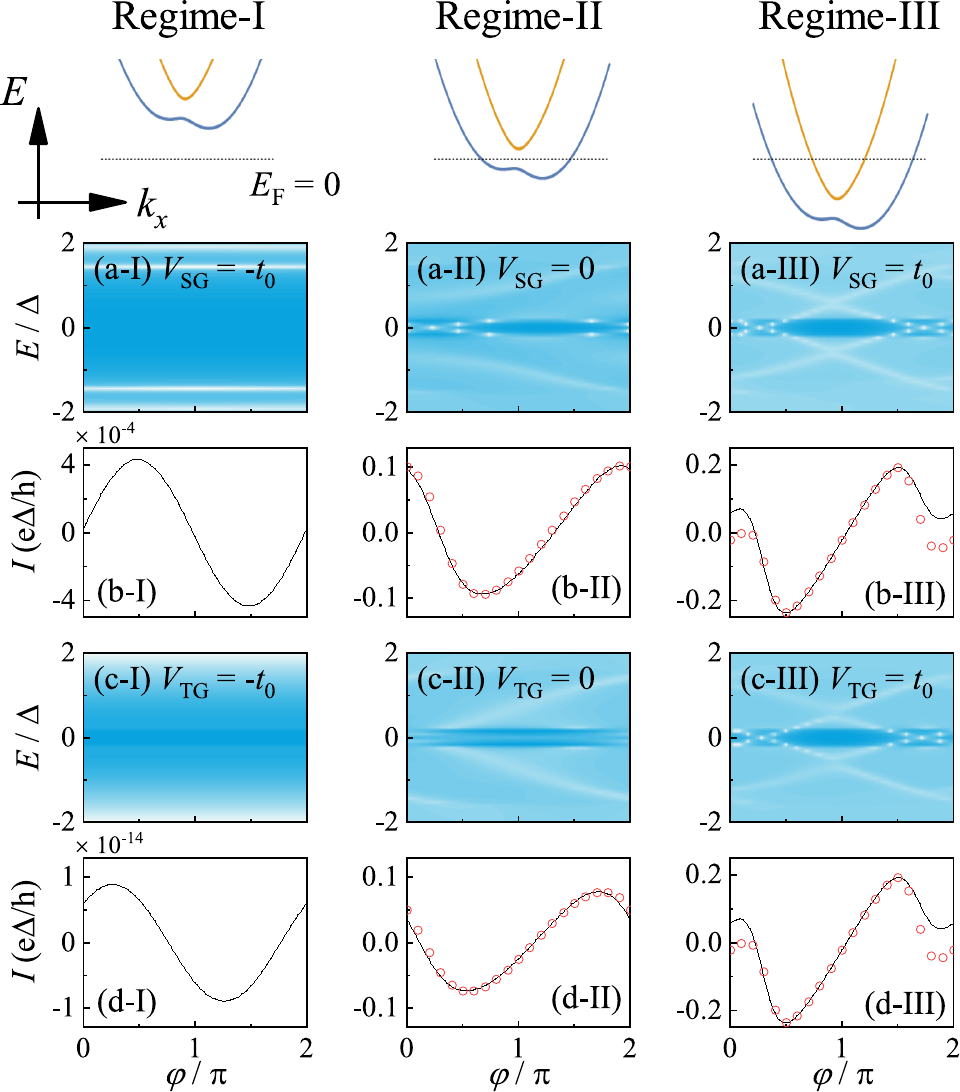}
\caption{The mechanism of the gate tunability of a Josephson diode of semiconductors. Top panel: the schematic dispersion of the dispersion of Eq. (\ref{h2deg}) with $t_0a^2=1$, $\alpha_0=1$, $\varphi_{so}=0.5\pi$, $m_y=0.1m_{0}$ and $m_{x/z}=0.2m_{0}$. The Fermi energy is $E_F=0$. The number of occupied bands is controlled by the voltage $V$. There are no, one, and two occupied bands in Regimes I, II, and III corresponding to $V_{SC}$ or $V_{TG}=-t_{0}$, $0$, and $+t_{0}$, respectively.
Panel (a) and (b) are the ABS and the CPR with $V_{TG}=t_{0}$ and $V_{SC}$ in Regime-I, II, and III.
Panel (c) and (d) are the ABS and the CPR with $V_{SC}=t_{0}$ and $V_{TG}$ in Regime-I, II, and III.
In panels (b) and (d), the red circles represent the current fit through Eq. (\ref{ifit}) with the parameters shown in Table \ref{tbalefit}.
}
\label{ABSCPREG}
\end{figure}
\begin{table}[t]
\caption{
Parameters of Eq. (\ref{ifit}) in Fig. (\ref{ABSCPREG}) where the critical current $I_c^\alpha$ is obtained by the ratio to the numerical results. }
\setlength{\tabcolsep}{3mm}{
\begin{tabular}{c|cccccc} 
\toprule
Fig.\ref{ABSCPREG} & $\bar{\tau}_{1}$ & $\bar{\tau}_{2}$ & $\varphi _{0}^{1}/\pi $ & $%
\varphi _{0}^{2}/\pi $ & $i_{D}^{1}$ & $i_{D}^{2}$ \\
\midrule
(b-II) & $0.9$ & $0.3$ & $0.7$ & $0.7$ & $0.0738$ & $0$ \\
(b-III) & $1$ & $1$ & $0.67$ & $-0.67$ & $-1$ & $1$ \\
(d-II) & $0.8$ & $0.3$ & $0.85$ & $0.85$ & $0.043$ & $0$ \\
(d-III) & $1$ & $1$ & $0.67$ & $-0.67$ & $-1$ & $1$\\
\bottomrule
\end{tabular}}
\label{tbalefit}
\end{table}
\begin{figure}[t]
\centering \includegraphics[width=0.48\textwidth]{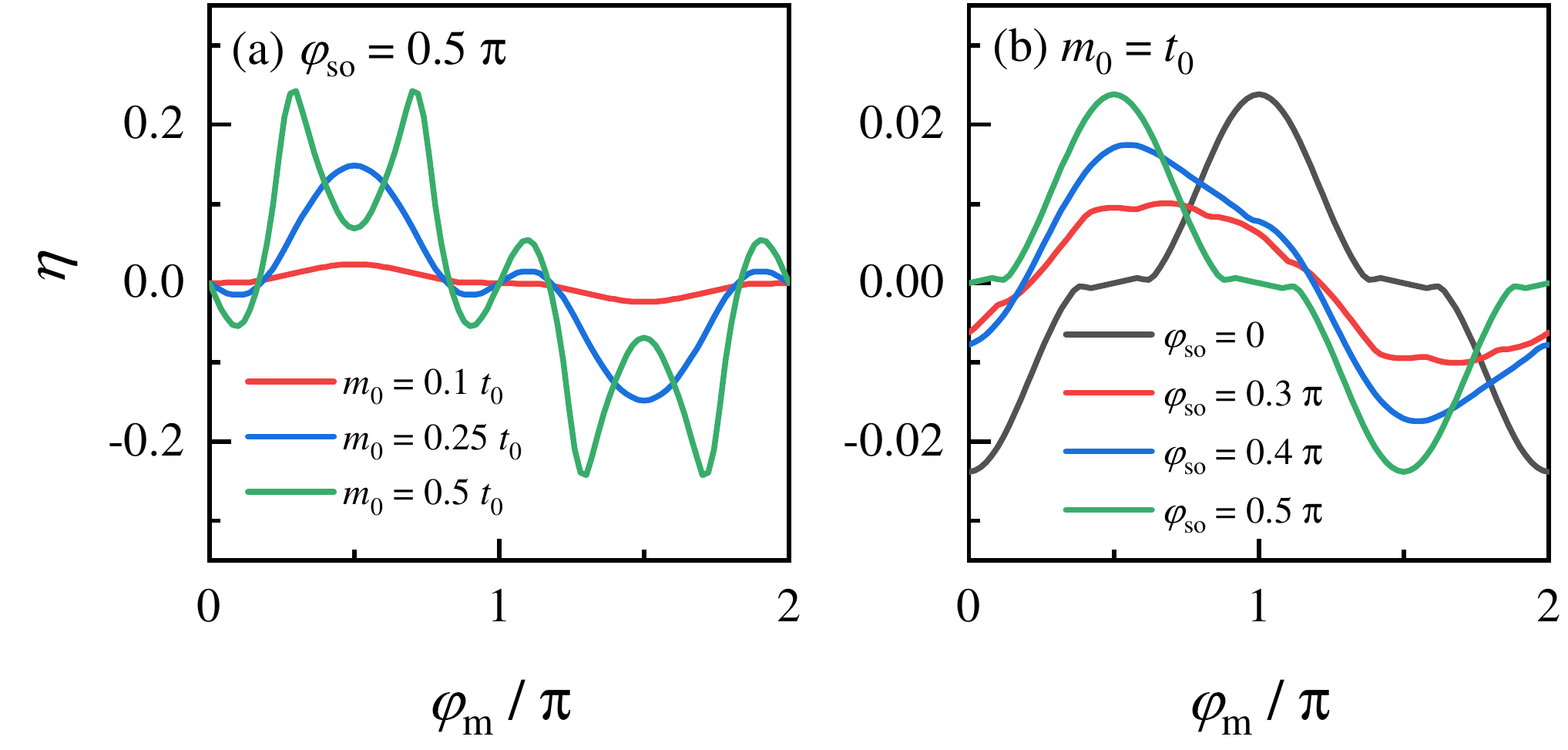}
\caption{The diode efficiencies of a 2D nanoflake with respect to the direction of the in-plane Zeeman field. (a) The case with pure RSOC ($\varphi_{SO}=0.5\pi$) with different field strengths. For the weak field, $\eta$ is nearly a sinusoidal function, $\eta \left( \varphi
_{m}\right) =\sum_{n}^{N}q_{n}\sin \left( n\varphi _{m}\right) $ with $%
q_{1}\gg q_{2}\gg \cdots \gg q_{N}$.
(b) The case with pure RSOC ($\varphi_{SO}=0.5\pi$), pure DSOC ($\varphi_{SO}=0$), and their mixture.
}
\label{etaphim}
\end{figure}

\subsubsection{Results and explanations}

Phenomenologically, there are two subband contributions to the total current, which can be expressed as $I_{1D}\left( \varphi \right) =\sum_{\alpha =1,2}I^{\alpha }\left(
\varphi \right) $ and 
\begin{eqnarray}
    I^{\alpha }\left( \varphi \right) /I_{c}^{\alpha }&=& i_{D}^{\alpha } + \frac{\sin \left( \varphi
+\varphi _{0}^{\alpha }\right) }{\sqrt{1-\bar{\tau}_{\alpha }\sin ^{2}\left( 
\frac{\varphi +\varphi _{0}^{\alpha }}{2}\right) }}\\
&&\times  \tanh \left[ \frac{\Delta(T) 
}{2k_{B}T}\sqrt{1-\bar{\tau}_{\alpha }\sin ^{2}\left( \frac{\varphi +\varphi
_{0}^{\alpha }}{2}\right) }\right]  \nonumber \label{ifit}
\end{eqnarray}
where $I_{c}^{\alpha }$, $\varphi _{0}^{\alpha }$, $\bar{\tau}_{\alpha }$, and $i_{D}^{\alpha }$ are the critical current, ground-state phase difference and the averaged transmission probability and the contribution from Doppler energy shift in the $\alpha ^{\text{th}}$ subband, respectively.
The parameters satisfying the fit with numerical results are shown in Table \ref{tbalefit}.
In Regime I of Fig. \ref{ABSCPREG} with $V_{SC}$, $V_{TG}\ll 0$, the system is an insulator and the low electron density fails to support Cooper-pair formation in the SC Region and propagating electron-hole pairs in the N Region, respectively.
Thus, no Josephson current and diode effect is allowed due to the low transmission probability and contribution from non-zero center-of-mass velocity ($\bar{\tau}_{\alpha }$, $i_{D}^{\alpha }\rightarrow 0$) as shown in Fig. \ref{ABSCPREG}(b-I) and (d-I).
No numerical fit is considered because this case is not important to our results. 

When $V_{SC}$, $V_{TG}$ are in Regime II, the system is half metal, where only one of the subbands is occupied resulting in one Josephson ballistic channel ($\bar{\tau}_{1}\sim 1$) with measurable contribution from non-zero center-of-mass velocity ($i_{D}^{1}\not=0$) while the other channel from the unoccupied subbands keeps in tunneling regime ($\bar{\tau}_{2}\ll 1$%
) and $i_{D}^{2}=0$. 
In both channels, there is a ground-state phase difference accumulated in the finite-length N region. 
This is justified in Fig. \ref{ABSCPREG}(b-II) and (d-II), where the numerical results can be fit by choosing suitable parameters in Eq. (\ref{ifit}).

Finally, when $V_{SC}$, $V_{TG}$ are in Regime III, both ballistic channels are activated. 
Since the interference between two channels is always constructive, and the ground-state phase differences are opposite, the diode efficiency oscillation is suppressed as shown in Fig. \ref{ABSCPREG}(b-III) and (d-III).
\subsubsection{Potential relation to the experiments}

Numerically, the current can be obtained through
\begin{equation}
J_{2D}\left( \varphi \right) =\sum_{k_{y}}J_{1D}\left( \varphi ,k_{y}\right) 
\end{equation}%
where $J_{1D}\left( k_{y}\right) $ can be calculated using the Green's function technique as in the 1D case with additional parameters relevant to $k_{y}=n_{y}\pi /W$ ($n_{y}$ is an integer). 
The diode efficiency is obtained in the CPR.

The dependence of diode efficiency on both the in-plane direction of the magnetic field and the gate voltage is exhibited in Fig. \ref{phimVSGVTG}, where the three regimes regarding zero-, one- and two-channel contributions controlled by the gates are expected, consistent with the discussions in 1D case. 
Besides, the diode efficiency oscillates as the TG varies.

Another feature of diode efficiency with respect to the magnetic field is $\eta \left( m_{y}\right) =-\eta \left( -m_{y}\right) $ with $m_{y}=m_{0}\sin \varphi _{m}$ as shown in Fig. \ref{etaphim}(a).
For a weak field intensity, the dependence of diode efficiency on $\varphi _{m}$ is not a simple sinusoidal function [$\eta \left( \varphi_{m}\right) \sim \sin \left( \varphi _{m}\right) $] as observed in Ref. \cite{Mazur2022}, but contains more higher-order terms $\eta \left( \varphi _{m}\right) =\sum_{n}^{N}q_{n}\sin \left( n\varphi _{m}\right) $ with $q_{1}\gg q_{2}\gg \cdots \gg q_{N}$, and only the first several orders dominate the contribution as the field intensity slightly increases. 
Furthermore, with a stronger field, the function of $\eta \left( \varphi _{m}\right) $ is intricate with a tendency as $\eta \left( \varphi _{m}\right) =X+\sin \left( X\right) $ and $X=\sum_{n}^{N}q_{n}\sin \left[ \left( 4n+1\right) \varphi _{m}\right] $.

Besides, in the weak-field situation, it is expected that the diode efficiency maximum occurs when $\varphi_{m}=\varphi _{so}$ as a signature of the alignment between the direction of the external magnetic field and the internal spin-orbit field. 
However, in our results, such a statement is only valid for a system with pure RSOI ($\varphi _{so}=\pi /2$) or pure DSOI ($\varphi _{so}=0$) as exhibited in Fig. \ref{etaphim}(b). 
In the system with max SOI, the maximum of $\eta \left( \varphi _{m}\right) $ do not occur at $\varphi_{m}=\varphi _{so}$\ and their behavior can not capture by a simple expression as well. 

\end{document}